\documentclass[12pt,a4paper]{article}
\usepackage{jheppub}
\usepackage{amsmath}
\usepackage{amsfonts}
\usepackage{amssymb}
\usepackage{graphicx}
\usepackage{mathrsfs}
\newcommand{\overbar}[1]{\mkern 1.5mu\overline{\mkern-4mu#1\mkern-4mu}\mkern 1.5mu}
\newcommand{\state}[1]{\left|#1\right>}
\newcommand{\phy}{\state{\textrm{phys}}}
\newcommand{\Tr}{\textrm{Tr}}
\newcommand{\tr}{\textrm{tr}}
\newcommand{\Det}{\textrm{Det}}
\newcommand{\nCr}[2]{\begin{pmatrix}#1 \\ #2\end{pmatrix}}
\usepackage[]{youngtab}

\makeatletter
\DeclareRobustCommand{\rvdots}{%
  \vbox{
    \baselineskip4\p@\lineskiplimit\z@
    \vskip0.1em
    \hbox{.}\hbox{.}
  }}
  \DeclareRobustCommand{\rhdots}{%
  \mbox{   
    \hskip0.1em
   $\cdot\cdot$
  }}
\makeatother

\newcommand{\pdfannotlink}{\pdfstartlink}

\def\Tiny{\fontsize{5pt}{5pt} \selectfont}

\title{Mixed symmetry Wilson-loop interactions in the worldline formalism}
\author[a]{James P. Edwards,}
\author[b,c]{Olindo Corradini}

\affiliation[a]{Department of Mathematical Sciences, \\ 
University of Bath\\
Claverton Down, Bath\\
BA2 7AY, UK}
\affiliation[b]{Dipartimento di Scienze Fisiche, Informatiche e Matematiche,\\ 
Universit\`a degli Studi di Modena e Reggio Emilia,\\ via Campi 213/A, I-41125 Modena, Italy} 
\affiliation[c]{INFN, Sezione di Bologna,\\ via Irnerio 46, I-40126 Bologna, Italy}

\emailAdd{jpe28@bath.ac.uk}
\emailAdd{olindo.corradini@unimore.it}

\abstract{Using the worldline formalism of the Dirac field with a non-Abelian gauge symmetry we show how to describe the matter field transforming in an arbitrary representation of the gauge group. Colour degrees of freedom are carried on the worldline by auxiliary fields, responsible for providing path ordering and the Wilson-loop coupling. The Hilbert space of these fields is reducible but we make use of recent work in order to project onto a single, arbitrary, irreducible representation. By functionally quantising the resulting theory we show that this procedure correctly generates the Wilson-loop interaction between the gauge field and the matter field taken to transform in a chosen representation. This work has direct application to physical observables such as scattering amplitudes in the presence of such a matter multiplet and lifts the restriction on the type of matter that has previously featured in worldline calculations.}

\begin{document}
\maketitle
\section{Introduction}
The worldline formalism \cite{Strass1, Schu} is a first quantised approach to quantum field theory which offers an alternative, highly efficient approach to many calculations. Recently it has been gaining in popularity as a valuable way of addressing modern problems in field theory, although it has roots in attempts to understand the Bern-Kosower master formulae \cite{Bern, Sato} which were originally uncovered in string theory. The general approach is to rewrite a field theory in terms of the quantum mechanics of a (spinning) point particle which traverses a particular worldline whilst interacting with a background gauge field. One then integrates over the particle trajectories and spin degrees of freedom to arrive at an effective action for the gauge field. This is often used as a starting point for the calculation of physically significant quantities and leads to simple and computationally economic calculations that maintain manifest gauge invariance~\cite{Sato, NaserTree}. For applications of this approach see, for example, the work in~\cite{Schmidt, Dunne, Bgrav1, hspin2, NaserTree, Anton, Bastbrst, Reuter, Mond, Bastianelli:2008vh, Bastianelli:2000dw}. 

Interest in worldline approaches to field theories with a non-Abelian symmetry has been revived by a new approach to the description of colour degrees of freedom. Traditionally one introduces a Lie-algebra valued potential which is minimally coupled to the matter field but gauge invariance demands that the path ordering procedure be implemented inside functional integrals. From the point of view of the worldline theory this is somewhat unnatural, for which reason it can be more beneficial to implement the coupling by introducing auxiliary worldline fields \cite{grass, grass2}. These additional colour fields may be Grassmann variables or may be bosonic in nature and are responsible for providing an enlarged Hilbert space and imposing the required path ordering \cite{Paul2, Me1}. However, this space is generally described by a reducible representation of the gauge group and it is necessary to extract a single wavefunction component which transforms in a desired irreducible representation \cite{Bastvt, Bastforms2}. In such work, one may choose the colour fields to transform in an arbitrary representation (and its conjugate) of any desired gauge group and it is subsequently possible to generate the Wilson-loop couplings for matter fields transforming in fully anti-symmetric or fully symmetric tensor products of this representation. In the present case we will instead achieve this coupling for an $SU(N)$ gauge group by combining sets of colour fields transforming only in the (conjugate-)\textit{fundamental} representation of this group; this approach, which we now describe, is easily generalised to other gauge groups with the same technique.

We have recently outlined the general method that achieves this projection onto an arbitrarily chosen representation of the symmetry group in \cite{JOdof}. To do this it was necessary to include a fixed number of families of the additional colour fields, each of which spans a Hilbert space whose wavefunction components transform in fully anti-symmetric or symmetric representations (depending on the nature of the auxiliary fields). The matter field then transforms in the (reducible) tensor product of these (anti-)symmetric representations but, by partially gauging a unitary symmetry which rotates between the families, we were able to impose constraints on the physical states in order to select a single irreducible representation from this space. This overcame previous limitations that had existed where the matter field was restricted to transforming in representations produced by fully anti-symmetric tensors product of the chosen representation of the colour fields \cite{Bastwl1}. In \cite{JOdof} we verified our construction by calculating the colour degrees of freedom associated to the matter field, confirming that this coincided with the dimension of the representation onto which we wished to project. 

In the current article we will go beyond this simple calculation to also include the coupling between the matter field and the gauge field. We will construct the worldline theory which describes the partition function of the Dirac field transforming in an arbitrary representation of the gauge group of $SU(N)$. We will carry out the functional quantisation of the colour degrees of freedom and show that it generates a sum over Wilson-loop interactions for particles in different representations of the symmetry group. We will then use the projection outlined in \cite{JOdof} to select from this sum the appropriate coupling for the field in the chosen irreducible representation. To begin with will choose each family of auxiliary fields to be anti-commuting so that everything is made up out of combinations of objects which transform in fully anti-symmetric representations. After achieving our aim we will then return to the case that the colour fields are bosonic, where the interactions are then constructed out of objects transforming in fully symmetric representations. 

In the next section we first present the standard reformulation of quantum field theory in a first quantised setting followed by the introduction of a single family of Grassmann colour carrying fields. In section \ref{secStruct} we discuss the structure of this worldline theory and its symmetries, which will allow us to arrive at the simplest form of projection based on gauging a $U(1)$ symmetry. We then quantise this basic model by computing the path integral on the circle in section \ref{secFunc}, with the result being the Wilson-loop interaction taken in an arbitrary fully anti-symmetric representation. The worldline theory is then extended in section \ref{secMixed}. There we introduce $F$ families of worldline fields and uncover a $U(F)$ symmetry on the worldline. We partially gauge this symmetry and compute the path integral to arrive at the Wilson-loop coupling for a particle in an arbitrarily chosen irreducible representation of the gauge group. Following this, section \ref{secSymm} briefly presents an analogous worldline theory in the case that the colour fields are bosonic, which is quantised in section \ref{secFuncB} and generalised to produce arbitrary irreducible representations in section \ref{secMixed2}.

\section{Worldline theory}
\label{SecTheory}
We begin with the configuration space action describing the $\mathcal{N} = 1$ supersymmetric point particle \cite{BdVH}:
\begin{equation}
	S\left[\omega, \psi\right] = \frac{1}{2}\int_{0}^{2\pi} d\tau \left[ \frac{\dot{\omega}^{2}}{T} + i\psi \cdot \dot{\psi}\right],
	\label{N=1}
\end{equation}
where $\omega^{\mu}$ are the bosonic coordinates of the particle in target space and the $\psi^{\mu}$ are their Grassmann super-partners encoding the particle's spin degrees of freedom. This action appears naturally in the worldline formalism \cite{Strass1, Strass2} of the Dirac field, $\Psi$, coupled to a gauge boson, $A$, which we now briefly review. The quantum effective action, $\Gamma\left[A\right]$, describes the dynamics of the gauge boson in the presence of the matter field. It is defined by integrating over the matter degrees of freedom which leads to a functional determinant:
\begin{align}
	\Gamma\left[A\right] &= -\ln{\left\{ \int \mathscr{D}(\bar{\Psi}\Psi) e^{-\int d^{4}x\, \bar{\Psi} \gamma \cdot D \Psi} \right\}  } \nonumber \\
	&= -\frac{1}{2} \Tr \ln{\left(\gamma \cdot D\right)^{2}}.
\end{align}
In the above we have denoted the covariant derivative by $D = \left(\partial + A\right)$, absorbing the coupling strength into the gauge field. Using the Schwinger proper time trick~\cite{Schwinger} the logarithm is turned into an integral, following which the functional trace is interpreted as the transition amplitude for a fictitious spin 1/2 point particle to traverse a closed loop in the presence of a background gauge field. This quantum mechanical transition amplitude can be written in path integral form as
\begin{equation}
	\int_{0}^{\infty} \frac{dT}{T} \oint \mathscr{D}(\omega, \psi) e^{-S_{E}\left[\omega, \psi\right]}~ \tr \mathscr{P}\exp{\left(     i\int_{0}^{2\pi} \mathcal{A}^{R}(\tau^{\prime})T^{R} d\tau^{\prime}\right)}.
	\label{worldline}
\end{equation}
where the action on the particle worldline is the (Euclidean) extension of (\ref{N=1}) to incorporate a (super-)Wilson loop interaction between the particle and the gauge field:
\begin{equation}
	g(\tau) := \mathscr{P}\exp{\left(i \int_{0}^{\tau} \mathcal{A}^{R}(\tau^{\prime})T^{R} d\tau^{\prime}\right)}; \qquad \mathcal{A} := \dot{\omega} \cdot A - \frac{iT}{2}\psi^{\mu}F_{\mu\nu}\psi^{\nu},
	\label{Wilson}
\end{equation}
where we have denoted the Hermitian Lie algebra generators by $\{T^{R}\}$ and indicated the path ordering prescription that is necessary for gauge invariance of the worldline theory. The colour degrees of freedom of the original Dirac field are now encoded in the above super-Wilson loop interaction between the fictitious particle and the background field: they are specified by choosing the representation of the $T^{R}$. In this way the original field theory based on second quantisation has been re-written in terms of simple one dimensional quantum mechanics which benefits from all of the advantages discussed in the introduction.

In the case that the gauge group is Abelian, the worldline theory enjoys translation invariance and a global supersymmetry. When the field strength tensor contains the additional commutator term present for a non-Abelian theory, the former of these invariances is preserved but the supersymmetry is spoilt at the level of the action. The traditional approach to preserve this supersymmetry is to incorporate only the Abelian part of $F_{\mu\nu}$ in the worldline action and to then introduce a modification to the path ordering prescription; this ``super-path ordering'' is responsible for completing the field strength tensor to include the missing commutator \cite{Schu} as discussed in Appendix~\ref{app:A}. However, as we shall now explain, there is an alternative approach which allows one to include the full field strength tensor in the worldline action whilst preserving the supersymmetry, at the expense of introducing additional, auxiliary worldline fields. 

The path ordering and the inclusion of the Lie algebra generators is unnatural from the perspective of the particle theory and tends to lead to unwanted complication when it comes to the calculation of physical quantities. It can instead be represented by introducing additional fields to encode the colour degrees of freedom \cite{Me1, Bastvt, Sam}. These fields can be anti-commuting or commuting as we have discussed in \cite{JOdof}. In this section we take the former route, using Grassmann auxiliary fields, leaving the bosonic case for section \ref{secSymm} onwards. For a gauge group $SU(N)$ we need $N$ pairs of fields $\tilde{\phi}^{r}$, $\phi_{r}$ which transform in the same (and conjugate) representation of the gauge group as $A$ (we follow the notation in \cite{Me1}). We choose the following Poisson brackets for these fields
\begin{equation}
	\{\tilde{\phi}^{r}, \phi_{s}\}_{PB} = -i\delta^{r}_{s}; \qquad \{\tilde{\phi}^{r}, \tilde{\phi^{s}}\}_{PB} = 0 = \{\phi_{r}, \phi_{s}\}_{PB}
	\label{phiPB}
\end{equation}
which can be used to absorb the gauge group indices of the generators by defining new objects which provide a representation of the Lie algebra
\begin{equation}
	R^{S} \equiv \tilde{\phi}^{r} (T^{S})_{r}{}^{s}\phi_{s};\qquad \left\{R^{S}, R^{T}\right\}_{PB}= f^{STU}R^{U},
	\label{Lie}
\end{equation}
where the $f^{STU}$ are the structure constants of the algebra. The action and the boundary conditions of these new variables is first order and is chosen so as to reproduce the path-ordering and trace in (\ref{worldline}). The extended worldline action including these ``colour fields'' is then given by
\begin{equation}
	 S[\omega, \psi, \tilde{\phi}, \phi] = \int_{0}^{2\pi} d\tau \left[ \frac{\dot{\omega}^{2}}{2T} + \frac{i}{2}\psi \cdot \dot{\psi} + i\tilde{\phi}^{r} \dot{\phi}_{r} + \tilde{\phi}^{r} \mathcal{A}_{r}{}^{s} \phi_{s}\right],
	 \label{worldlinephi}
\end{equation}
where we have introduced the notation $\mathcal{A}_{r}{}^{s} = \mathcal{A}^{R}(T^{R})_{r}{}^{s}$. As promised, this action now enjoys invariance under a global supersymmetric transformations generated by the constant Grassmann parameter $\eta$, in which the worldline fields transform as
\begin{alignat}{5}
	\delta_{\eta}\omega &= i\eta \psi; && \delta_{\eta} \psi &&= -\frac{\eta}{T} \dot{\omega} \nonumber \\
	 \delta_{\eta} \tilde{\phi}^{r} &= -\eta \tilde{\phi}^{s}\psi \cdot A_{s}{}^{r}; \qquad &&\delta_{\eta} \phi_{r} &&= -\eta \psi \cdot A_{r}{}^{s}\phi_{s}
	 \label{susyphi}
\end{alignat}
We will gauge these symmetries in the next section where the theory is reformulated in superspace so as to arrive at a worldline theory with local supersymmetry and reparameterisation invariance. This is useful for a deeper understanding of the structure of the worldline theory. The particle action and global symmetry discussed above will then arise upon a certain choice of gauge fixing, but this approach will also require constraints to be imposed on the physical state space. Note also that the worldline Green function for the $\tilde{\phi}$, $\phi$ theory is (up to constants which depend upon the boundary conditions) proportional to the step function $2G(\tau, \tau^{\prime}) = \theta(\tau - \tau^{\prime})$, which suffices to build the (familiar) path ordering -- further details are given in the next section and in Appendix~\ref{app:A}.

We have studied this action before in the context of the standard model and unified theories \cite{Paul2, Me1} and it has also appeared in perturbative calculations in the worldline formalism \cite{Bastwl1, Bastwl2}. Related actions include the generalisation of the model to $O(2S)$ extended supersymmetry \cite{hspin, hspin3, Corradini:2010ia}, whereby $2S$ families of Grassmann fields $\{\psi_{k}\}$ are used to describe particles of (semi-)integer spin $S$, in flat and curved spaces. We shall make use of some of these ideas in later sections. For now we focus on the action (\ref{worldlinephi}), the simplest incorporation of colour fields into the worldline model. In the next section we consider the structure of this theory by exploring the Hilbert space of physical states using canonical techniques. We will also discuss the gauging of the reparameterisation invariance and supersymmetry. This will help us to interpret the results of the functional quantisation presented in section \ref{secFunc}. These results will then be recycled in the remaining parts of this article when we generalise the model to describe non-trivial matter multiplets.

\section{Worldline symmetry and canonical structure}
\label{secStruct}
The supersymmetry of the $x^{\mu}, \psi^{\mu}$ theory can be extended to the theory describing $\tilde{\phi}$ and $\phi$ if we introduce two commuting auxiliary variables $\tilde{z}$ and $z$ to be their super-partners and replace the part of the action involving these fields in (\ref{worldlinephi}) by
\begin{align}
	S[\tilde{\phi}, \tilde{z} ; \phi, z] =  \int d\tau \left[\tilde{\phi}^{r}\left(i\delta_{r}^{s}\frac{d}{d\tau} + \mathcal{A}^{0}_{r}{}^{s}\right)\phi_{s}  -T\tilde{z}^{r}z_{r} + iT\left(\tilde{z}^{r}\psi \cdot A_{r}{}^{s} \phi_{s} + \tilde{\phi}^{r} \psi \cdot A_{r}{}^{s} z_{s}\right)\right],
	\label{zz}
\end{align}
where $\mathcal{A}^{0} = \dot{\omega}\cdot A  -iT\psi \cdot \partial A \cdot \psi$ is the super-Wilson loop exponent without the commutator part of the field strength tensor. It is easy to verify that the full Wilson loop is now generated (i.e $\mathcal{A}^{0}$ is completed to include the non-Abelian field strength tensor) by integrating over the auxiliary fields $\tilde{z}$ and $z$ \cite{Aux, JOdof} and that this action is also globally supersymmetric if
\begin{align} \delta_{\eta}\tilde{\phi} &= i\eta \tilde{z}; \qquad \delta_{\eta}\tilde{z} = -\frac{\eta}{T}\dot{\tilde{\phi}} \nonumber \\
\delta_{\eta}\phi &= i\eta z,\qquad \delta_{\eta}z= -\frac{\eta}{T}\dot\phi
\label{susycz}
\end{align} 
for the constant Grassmann generator $\eta$. Note that when the fields $\tilde{z}$ and $z$ are put on-shell the transformations for $\tilde{\phi}$ and $\phi$ given in (\ref{susyphi}) are recovered. Indeed, since these fields are auxiliary, we may trivially integrate them out, in which case we recover the action (\ref{worldlinephi}). 

This global supersymmetry is the residual symmetry which comes from gauge fixing a locally supersymmetric action, which we give here in the superspace formalism\footnote{Local supersymmetry transformations provide a graded generalisation of the diffeomorphisms of the circle, $\textrm{Sdiff}\left(S^{1}\right)$ -- see \cite{West}. Under transformations parameterised by $V\left(\tau\right)$, the generator of reparameterisations, and a Grassmann function $\eta\left(\tau\right)$, generating pure supersymmetry transformations,
\begin{equation}
\tau \rightarrow \tau + V\left(\tau\right) + \theta \eta\left(\tau\right); \qquad \theta \rightarrow \theta + \eta\left(\tau\right) + \frac{1}{2}\theta \dot{V}\left(\tau\right),
\end{equation}
the super-derivative transforms homogeneously
\begin{equation}
	D \mathbf{X} \rightarrow \Lambda\left(\tau, \theta\right) D \mathbf{X}
\end{equation}
and the super-einbein transforms as
\begin{equation}
	\mathbf{E} \rightarrow \Lambda^{2}\left(\tau, \theta\right) \mathbf{E}
\end{equation}
where $\Lambda\left(\tau, \theta\right) = 1 + \frac{1}{2}\dot{V}\left(\tau\right) + \theta \dot{\eta}\left(\tau\right)$. The integration measure transforms as $d\tau d\theta \rightarrow \Lambda^{-1}\left(\tau, \theta\right) d\tau d\theta$, ensuring that $\mathbf{E}d\tau d\theta$ transforms homogeneously;  $\tilde{\mathbf{\Phi}}$ and $\mathbf{\Phi}$ transform as worldline scalars like $\mathbf{X}$.\label{footSusy}}. We extend the parameter domain $\tau \rightarrow (\tau, \theta)$, where the Grassmann parameter $\theta$ squares to zero, and introduce an einbein, $e\left(\tau\right)$, and its super-partner (the gravitino), $\chi\left(\tau\right)$. We also define the superfields
\begin{align}
	\mathbf{X}\left(\tau, \theta\right) &= \omega\left(\tau\right) + \theta e^{\frac{1}{2}}\left(\tau\right)\psi\left(\tau\right), \nonumber \\
	\mathbf{E}\left(\tau,\theta\right) &= e\left(\tau\right) - 2\theta e^{\frac{1}{2}}\left(\tau\right)\chi\left(\tau\right), \nonumber \\
	\tilde{\mathbf{\Phi}}\left(\tau, \theta\right) &= \tilde{\phi}\left(\tau\right) + \theta e^{\frac{1}{2}}\left(\tau\right)\tilde{z}\left(\tau\right), \nonumber\\
	\mathbf{\Phi}\left(\tau, \theta\right) &= \phi\left(\tau\right) + \theta e^{\frac{1}{2}}\left(\tau\right)z\left(\tau\right),
\end{align}
and the super-derivative $D = \partial_{\theta} + i\theta \partial_{\tau}$. We have suppressed the colour indices on $\tilde{\mathbf{\Phi}}$ and $\mathbf{\Phi}$ for brevity and chosen a convention for each superfield which leads to an action which takes a similar form to that of Brink, diVecchia and Howe \cite{BdVH}. Then the locally supersymmetric action
\begin{equation}
	\int\! d\tau d\theta \left[ -\frac{1}{2}\mathbf{E}^{-1} D^{2}\mathbf{X} \cdot D\mathbf{X} - \tilde{\mathbf{\Phi}}^{r}D\mathbf{\Phi}_{r} + i\tilde{\mathbf{\Phi}}^{r} D \mathbf{X} \cdot A_{r}{}^{s}\left(\mathbf{X}\right) \mathbf{\Phi}_{s}\right],
\end{equation}
can be expanded into component fields and integrated over $\theta$ to give the configuration space action
\begin{align}
	\int d\tau \bigg[ \frac{1}{2}e^{-1}\dot{\omega}^{2} + \frac{i}{2}\psi \cdot \dot{\psi} + i\tilde{\phi}^{r}  \dot{\phi}_{r} &- \frac{i\chi}{e}\dot{\omega} \cdot \psi + \tilde{\phi}^{r} \mathcal{A}^{0}_{r}{}^{s} \phi_{s}  \nonumber \\
	 &- e \tilde{z}^{r}{z}_{r} + ie(\tilde{z}^{r} \psi \cdot A_{r}{}^{s} \phi_{s} + \tilde{\phi}^{r} \psi \cdot A_{r}{}^{s} z_{s})\bigg]
	\label{cpt}
\end{align}
The transformations of the components are inherited from the superspace variations in footnote \ref{footSusy}. Under reparameterisations $\omega$, $\psi$, $\tilde{\phi}$, $\tilde{z}$, $\phi$ and $z$ transform as worldline scalars, $e^{2}$ transforms as a worldline metric and $\chi$ transforms as $e$. Under a pure supersymmetry transformation generated by the Grassmann $\eta(\tau)$ the transformations of the fields are
\begin{equation}
	\delta_{\eta} \omega = i\eta \psi; \quad \delta_{\eta} \psi = -\frac{\eta}{e}\left( \dot{\omega} - i \chi \psi\right); \quad \delta_{\eta}e = -2i\eta \chi;\quad \delta_{\eta} \chi = \dot{\eta},
	\label{susy}
\end{equation}
and similarly
\begin{align}
	\delta_{\eta} \tilde{\phi} = i\eta \tilde{z}; \quad \delta_{\eta} \tilde{z} = -\frac{\eta}{e}\left( \dot{\tilde{\phi}} -    i\chi \tilde{z}\right); \quad \delta_{\eta} \phi = i\eta z; \quad \delta_{\eta} z = -\frac{\eta}{e}\left( \dot{\phi} -    i\chi z\right),
	\label{susy2}
\end{align}
and the action (\ref{cpt}) is invariant under these variations.
 
With the given periodicity conditions on the worldline fields the local supersymmetry can be gauged to $e(\tau) = T$ and $\chi = 0$ where $T$ is a constant modulus (we discuss the Faddeev-Popov determinant in the next section) -- we note that in the case of open worldlines the gauge fixing procedure can only fix $\chi$ to be a constant Grassmann number, $\chi_{0}$, say, which is another modulus to integrate over. We intend to study this case in future work. Substituting their classical solutions into (\ref{cpt}) reproduces the action for both the matter fields $S\left[\omega, \psi\right]$ and the new Grassmann fields $S[\tilde{\phi}, \phi]$ upon integration over $\theta$. The equations of motion for these variables impose first class constraints\footnote{The equation of motion for the einbein enforces the mass-shell condition on physical states and that of the gravitino corresponds to the Dirac equation $\gamma \cdot D \phy$ = 0, when the fields $\psi^{\mu}$ are replaced by $\gamma$-matrices upon quantisation.} on the resulting theory \cite{BdVH}. One of the great advantages that the superspace form of the theory offers is that the gauge field $A$ now enters linearly in the action, making its functional quantisation straightforward. In Appendix~\ref{app:A} we discuss different approaches to producing the path ordered exponential $g(2\pi)$ out of this locally supersymmetric theory by using the form of the Green functions of the Grassmann variables to construct a super-path ordering and in the next subsection we examine the Hilbert space of the $\tilde{\phi}$, $\phi$ theory.

\subsection{Fock space of the colour fields}
\label{Fock}
We must also briefly discuss the Hilbert space of the extended theory so as to understand the form of the results we present in the next sections. We follow the procedure presented in \cite{JOdof}. For canonical quantisation we promote $\tilde{\phi}^{r}$ and $\phi_{s}$ to creation and annihilation operators which span a two dimensional Fock space $\{\state{\downarrow_{r}}, \state{\uparrow_{r}}\}$ for each index $r$. The Poisson brackets (\ref{phiPB}) become the fundamental anti-commutation relations, $\{\hat{\phi}^{\dagger r}, \hat{\phi}_{s}\} = \delta^{r}_{s}$, which in a coherent state basis are solved by setting $\hat{\phi}^{\dagger r} = \tilde{\phi}^{r}$ and $\hat{\phi}_{r} = \partial_{\tilde{\phi}^{r}}$ when acting on wave-functions $\Psi(x, \tilde{\phi})$. The Fock space is then built up by acting on the vacuum with creation operators and is finite dimensional since we have taken $\tilde{\phi}$ and $\phi$ to anti-commute. The creation operators $\tilde{\phi}^{r}$ have an index which transforms in the conjugate representation of the generators $T^{R}$ (see (\ref{Lie})) and the wave function $\Psi(x, \tilde{\phi})$ has a finite expansion in components which transform in anti-symmetric products of this representation \cite{Howe}:
\begin{equation}
	\Psi(x, \tilde{\phi}) = \Psi(x) + \tilde{\phi}^{r_{1}} \Psi_{r_{1}}(x) + \tilde{\phi}^{r_{1}}\tilde{\phi}^{r_{2}} \Psi_{r_{1}r_{2}}(x) + \ldots + \tilde{\phi}^{r_{1}}\tilde{\phi}^{r_{2}}...\tilde{\phi}^{r_{N}}\Psi_{r_{1}r_{2}...r_{N}} ,
\end{equation}
where the wavefunction components transform in the fully anti-symmetric representations of the gauge group so that $\Psi(x, \tilde{\phi})$ will in general be described by a reducible representation of the gauge group \cite{Hoker}. 

To pick out a given representation requires a means of projecting the intermediate states of the path integral onto coherent states with the correct occupation number \cite{Hoker}. The operator $\hat{n} = \hat{\phi}^{\dagger r}\phi_{r}$ has the commutator $[\hat{n}, \hat{\phi}^{\dagger r}] = \phi^{\dagger r}$ and therefore has eigenstates $\hat{n}\state{\downarrow_{r}} = 0,\, \hat{n}\state{\uparrow_{r}} = \state{\uparrow_{r}}$ whose eigenvalues indicate the occupation of state $r$. So we can impose the constraint that the physical states have occupation number $n$ via the introduction of a delta function in the path integral measure:
\begin{equation}
	\int \mathscr{D}\tilde{\phi}\mathscr{D}\phi\ \rightarrow \int \mathscr{D}\tilde{\phi}\mathscr{D}\phi\, \delta\big(\tilde{\phi}^{r}\phi_{r} - (n -\frac{d_{R}}{2})\big)
\end{equation}
where $d_{R}$ is the dimension of the representation in which the fields transform. On the Fock space (anti-symmetrising to resolve the operator ordering ambiguity) this constraint becomes
\begin{equation}
\left(\frac{1}{2}\big(\hat{\phi}^{\dagger r} \hat{\phi}_{r} - \hat{\phi}_{r} \hat{\phi}^{\dagger r}\big) -  (n -\frac{d_{R}}{2})\right)\phy= 0.
\end{equation}
This acts on the wave function as $\left(\tilde{\phi}^{r} \partial_{\tilde{\phi}^{r}} - n\right)\Psi = 0$ which is easily seen to enforce the vanishing of all components of $\Psi$ except for that which transforms under the representation with $n$ indices. In functional quantisation we can represent the delta function as a path integral over a further worldline field so as to form
\begin{equation}
	\int \mathscr{D}\tilde{\phi}\mathscr{D}\phi \mathscr{D} a\, e^{ i\int d\tau \left[ (n -\frac{d_{R}}{2}) a(\tau) -  a(\tau) \tilde{\phi}^{r} \phi_{r}\right]}.
	\label{inta}
\end{equation}
In this way the field $a$ acts as a Lagrange multiplier whose equation of motion imposes the correct constraint on physical states.

An equivalent way of arriving at (\ref{inta}) -- and one which will be important in later sections -- is to recognise that the action (\ref{worldlinephi}) has a global symmetry under rotations of $\tilde{\phi}$ and $\phi$ by a $U(1)$ phase factor: under $\phi_{r} \rightarrow e^{-i\vartheta}\phi_{r}$ and $\tilde{\phi}^{r} \rightarrow \tilde{\phi}^{r}e^{i\vartheta}$ the action remains invariant. The conserved current associated to this transformation is easily verified to be $\tilde{\phi}^{r} \phi_{r}$ which upon quantisation just becomes the occupation number operator. To gauge this symmetry we can introduce a new worldline field $a(\tau)$ and construct a covariant derivative $D\phi = \left(\frac{d}{d\tau} + i a\right)\phi$. The ``dynamics'' of this gauge field is given by a Chern-Simons term  \cite{Bastwl1, Bastwl2, Bastvt} which on $S^{1}$ can only be the one-form $\int d\tau \,(n -\frac{d_{R}}{2}) a(\tau)$. Note that the quantised nature of the prefactor multiplying $a(\tau)$ ensures that this Chern-Simons theory is absent of anomalies \cite{anom}. We hence arrive at (\ref{inta}) which is the gauged version of (\ref{worldlinephi}). Using the covariant derivative we can return to the worldline theory to write the action corresponding to the $\tilde{\phi}$, $\phi$ part of (\ref{worldlinephi}) as
\begin{equation}
 S[\tilde{\phi}, \phi, a] = \int_{0}^{2\pi}d\tau \left[\tilde{\phi}^{r} \left(i\delta_{rs} D + \mathcal{A}_{r}{}^{s}\right)\phi_{s} +(n -\frac{d_{R}}{2})\, a\right].
\label{Za'}
\end{equation}
The Grassmann fields are taken to have anti-periodic boundary conditions whilst the gauge field is periodic on the interval. The infinitesimal transformations of the fields under the $U(1)$ symmetry which leave this action invariant are generated by $\vartheta(\tau)$ and are given by
\begin{equation}
	\delta_{\vartheta}\tilde{\phi}^{r} = i\vartheta\tilde{\phi}^{r}; \qquad \delta_{\vartheta}\phi_{r} = -i\vartheta\phi_{r}; \qquad \delta_{\vartheta}a =\dot{\vartheta}
\end{equation}
This mechanism has been used very successfully in the worldline formalism of quantum field theory for a variety of situations \cite{Dai, Bastforms, Bastbrst, Bastvt, Bastwl2, Howe} in order to project onto irreducible representations of the gauge group. In \cite{Paul2, Me1}, this projection was not included so as to count the contribution from all representations constructed out of totally anti-symmetric tensor products of the fundamental representations of $SU(2)$, $SU(3)$, $SU(5)$ and $SO(10)$ in order to investigate worldline approaches to the standard model and unified theory.

In this article we include the $U(1)$ gauge field and demonstrate how the projection onto irreducible representations works in the functional approach. These results will be the building blocks of later sections, where we will extend the worldline theory of this section to include different families of Grassmann variables to construct worldline theories of matter fields that transform in arbitrary irreducible representations of the gauge group. Once this has been attained, we will also explain how it is possible to take the fields $\tilde{\phi}$ and $\phi$ to be bosonic in order to generate an infinite sum over fully symmetric representations, as in \cite{Bastwl2}.

\section{Functional Quantisation}
\label{secFunc}
In this section we carry out the (Euclidean) path integral quantisation of the gauged worldline theory discussed above. In order to do so we will follow the usual procedure of choosing a convenient fixing of the gauge symmetry, introducing Faddeev-Popov determinants to compensate for this restriction and integrating over the remaining degrees of freedom. This will be seen to compute the exponentiated line integrals in various representations of the $SU(N)$ gauge symmetry, of which the $U(1)$ projection outlined above will select one irreducible representation. We continue for the time being by assuming that $\tilde{\phi}$ and $\phi$ are anti-commuting (Grassmann) fields, discussing what happens if they are taken to be commuting in later sections. The results of this section will be vital when we come to generalise the model to include multiple copies of the colour fields so as to project on to an arbitrary representation of the gauge group.

We have three local symmetries to fix -- namely the reparameterisation invariance, the supersymmetry and the $U(1)$ invariance. These are gauged by the triple of fields $(e,\,\chi,\, a)$. We gauge fix these to constants $(T,\, 0,\, \frac{\theta}{2\pi})$ which is particularly convenient from the point of view of calculations and is standard in the worldline approach to quantum field theory \cite{hspin, Bastwl1}. These moduli parameterise gauge inequivalent choices of the fields. It is well known that the Faddeev-Popov measure associated to fixing the einbein is $\frac{dT}{T}$ and on closed curves the anti-periodicity of $\chi$ allows us to completely gauge it away without requiring the introduction of ghosts (for further details see the appendices of \cite{Us2}). For the fixing of the $U(1)$ field we discuss in Appendix~\ref{app:B} that we can at best set the gauge field equal to a constant and application of large gauge transformations implies that $\theta$ must be interpreted as an angle in $[0, 2\pi]$. The measure associated with this gauge fixing is independent of $\theta$ and can be normalised to $\frac{d\theta}{2\pi}$. In summary, for an arbitrary functional $\Omega$ we can make the replacement
\begin{equation}
	\int\mathscr{D}e\mathscr{D}\chi \mathscr{D}a\, \Omega\left[e(\tau), \chi(\tau), a(\tau)\right] \rightarrow \int_{0}^{\infty}\frac{dT}{T}\int_{0}^{2\pi} \frac{d\theta}{2\pi}\, \Omega\left[T, 0, \theta\right]
	\label{measures}
\end{equation}
which will be used to put path integrals into gauge fixed form.

Using (\ref{measures}), Wick rotating to Euclidean space and recalling that $\tilde{z}$ and $z$ are non-dynamical auxiliary fields that can be integrated out at any time we will consider the partition function
\begin{align}
	\mathcal{Z}\left[\mathcal{A}\right] \!&=\! \int\frac{\mathscr{D}\omega \mathscr{D}\psi \mathscr{D}\tilde{\phi} \mathscr{D}\phi \mathscr{D}e\mathscr{D}\chi \mathscr{D}a}{\textrm{Vol(Gauge)}}\, e^{- \int_{0}^{2\pi} d\tau \left[ e^{-1}\frac{\dot{\omega}^{2}}{2} + \frac{1}{2}\psi \cdot \dot{\psi} - \frac{\chi}{e}\dot{\omega} \cdot \psi + \tilde{\phi}^{r}\left(\delta_{r}^{s}D -i \mathcal{A}_{r}{}^{s} \right)\phi_{s} - i(n - \frac{d_{R}}{2})a \right]} \nonumber \\
	&= \!\int_{0}^{\infty} \frac{dT}{T} \oint \mathscr{D}\omega \mathscr{D}\psi\, e^{-\frac{1}{2} \int_{0}^{2\pi} \frac{\dot{\omega}^{2}} {T} + \psi \cdot \dot{\psi} \, d\tau}\int_{0}^{2\pi} \frac{d\theta}{2\pi}\, e^{i(n -\frac{d_{R}}{2}) \theta } \mathcal{Z}\left[\mathcal{A}, \theta\right].
	\label{Z}
\end{align}
We have denoted by $\mathcal{Z}\left[\mathcal{A}, \theta\right]$ the partition function of the $\tilde{\phi}$, $\phi$ theory which is responsible for generating the Wilson-loop:
\begin{equation}
	\mathcal{Z}\left[\mathcal{A}, \theta\right] = \int\mathscr{D}\tilde{\phi} \mathscr{D}\phi\, e^{- \int_{0}^{2\pi} d\tau\, \tilde{\phi}^{r}(\delta_{r}^{s}D -i \mathcal{A}_{r}{}^{s}) \phi_{s}},
	\label{Za}
\end{equation}
where $\mathcal{A}$ is the super-Wilson loop exponent which, on the chosen gauge slice, depends on the path $\omega$, spin variable $\psi$ and modulus $T$. In the first line of (\ref{Z}) we divided by the size of the symmetry group associated to the reparameterisation invariance, the supersymmetry and the $U(1)$ symmetry, which is taken care of in the second line by the integration over the moduli $T$ and $\theta$. The covariant derivative becomes $D\phi = (\frac{d}{d\tau} + \frac{i\theta}{2\pi})\phi$ after gauge fixing. It is quite illuminating to see the coupling between the Grassmann fields in (\ref{Za}) which consists of a piece $-i\mathcal{A}_{rs}$ representing the particle's interaction with the space-time gauge field and a second coupling to the worldline gauge field $\frac{i\theta}{2\pi} \delta^{s}_{r}$ whose interplay will be crucial in the proceeding sections.  

The integral over the physical worldline fields $\omega$ and $\psi$ in (\ref{Z}) is the usual worldline integral that arises in the worldline formalism of quantum field theory. The path integral of these variables coupled to Wilson loops is well established in the literature \cite{Schu} and is not the source of any novelty in this article. We will instead focus on $\mathcal{Z}\left[\mathcal{A}, \theta\right]$, calculating the gauge group information that it contains, before completing the integral over $\theta$ in order to select out the path ordered exponent in a chosen anti-symmetric representation. 

For the purposes of one-loop calculations in the worldline approach it is convenient to incorporate the coupling to the gauge field into the Grassmann fields by a re-definition \cite{Bastwl1, Bastwl2} 
\begin{align}
	\tilde{\phi}\left(\tau\right) &\rightarrow \tilde{\phi}\left(\tau\right)\exp{\left(-i \int_{0}^{\tau} a\left(\tau^{\prime}\right)d\tau^{\prime}\right)} = \tilde{\phi}(\tau)e^{- \frac{i\theta}{2\pi} \tau}\nonumber \\
	 \phi\left(\tau\right) &\rightarrow \exp{\left(i \int_{0}^{\tau} a\left(\tau^{\prime}\right)d\tau^{\prime}\right)}\phi\left(\tau\right) = e^{\frac{i\theta}{2\pi} \tau}\phi(\tau)
	 \label{redef}
\end{align}
whose effect is to remove the dependence on $a$ in the Lagrangian of (\ref{Za}) at the expense of changing the boundary conditions on the Grassmann fields. The original theory initially required $\tilde{\phi}$ and $\phi$ to be anti-periodic so that under the above redefinition the boundary conditions become $\tilde{\phi}\left(2\pi\right) = -e^{i\theta}\tilde{\phi}\left(0\right)$ and $\phi\left(2\pi\right) = - e^{-i\theta}\phi(0)$. The integral over $\theta$ in (\ref{Z}) then interpolates between all such ``twisted'' boundary conditions. This procedure is particularly useful for perturbative calculations because it decouples the Grassmann fields from the worldline, or $U(1)$, gauge field. In this context the Green function of the Grassmann fields is modified to reflect the field re-definition\footnote{As has been discussed earlier the Green function with anti-periodic boundary conditions is $G(\tau, \tau^{\prime}) = \frac{1}{2}\left(\Theta(\tau - \tau^{\prime}) - \Theta(\tau^{\prime} - \tau)\right)$. With the twisted boundary conditions above this is replaced by $G_{\theta}(\tau, \tau^{\prime}) = \frac{1}{2\cos{\frac{\theta}{2}} }\left(e^{i\frac{\theta}{2}}\Theta(\tau - \tau^{\prime}) - e^{-i\frac{\theta}{2}}\Theta(\tau^{\prime} - \tau)\right)$, which clearly reduces to the original Green function for $\theta = 0$. It should be noticed that this Green function may appear with derivatives,
for example when performing integrations by parts on the coordinates
$\omega^\mu(\tau)$.
Eventual singularities (derivative of the step functions) cancel using
standard regularisations of the path integral which preserve gauge invariance.
In curved space one needs instead explicit counterterms to preserve background
local symmetries, see for example the general results of~\cite{Bastianelli:2011cc}, needed
in treating higher spinning particles in curved space~\cite{Bastianelli:2012bn, Bastianelli:2014lia}.}. We will also use this procedure for the calculation of $\mathcal{Z}\left[\mathcal{A}, \theta\right]$ (we comment below on how our calculation would differ if we did not make the field redefinition).

With this change of variables the partition function becomes
\begin{equation}
	\mathcal{Z}\left[\mathcal{A}, \theta\right] = \int_{\textrm{TBC}}\mathscr{D}\tilde{\phi} \mathscr{D}\phi\, e^{- \int_{0}^{2\pi} d\tau\, \tilde{\phi}^{r}(\delta_{r}^{s}\frac{d}{d\tau} -i \mathcal{A}_{r}{}^{s}) \phi_{s}},
	\label{Zat}
\end{equation}
where $\mathrm{TBC}$ stands for the twisted boundary conditions on $\tilde{\phi}$ and $\phi$. Integrating over $\bar{\phi}$ and $\phi$ in (\ref{Zat}) leads to a functional determinant 
\begin{align}
	\mathcal{Z}\left[\mathcal{A}, \theta\right] &= \underset{\scriptscriptstyle\rm TBC}{\det}{\left(i\left(\frac{d}{d\tau} -i \mathcal{A}\right)\right)},
	\label{Zdet}
	\end{align}
which we evaluate in the next section as in \cite{Paul2, Me1}.
\subsection{Calculation of the determinant}	
We define the functional determinant (\ref{Zdet}) as the product of the eigenvalues of the operator in brackets on the space of fields satisfying the twisted boundary conditions. To find the eigenfunctions of this operator, $v(\tau)$, we make use of the defining equation of the super-Wilson loop, (\ref{Wilson}), writing them as $v\left(\tau\right) = g\left(\tau\right)f\left(\tau\right)$. Then the eigenvalue equation $i\left(\frac{d}{d\tau} + \mathcal{A}\right)v\left(\tau\right) = \mu v\left(\tau\right)$ translates to an equation for $f\left(\tau\right)$:
\begin{equation}
	i \frac{d}{d\tau}f\left(\tau\right) = \mu f\left(\tau\right) \Longrightarrow f\left(\tau\right) = v\left(0\right)e^{-i \mu \tau}.
\end{equation}
We then impose the twisted boundary conditions which allows us to relate the eigenvalues $\mu$ to the eigenvalues of the Wilson loop, $\rho$. Indeed, taking $v\left(2\pi\right) = -e^{-i\theta}v\left(0\right)$ we require $v\left(0\right)$ to be an eigenvector of $g\left(2\pi\right)$ and must impose a condition on $\mu$:
\begin{equation}
	g\left(2\pi\right)v\left(0\right) = \rho v\left(0\right); \qquad \rho e^{-2\pi i \mu} = -e^{-i \theta}.
	\label{v0}
\end{equation}
These equations can be solved to write the eigenvalues $\mu$ in terms of the eigenvalues of the Wilson-loop: $\mu = n + \frac{1}{2} + \frac{\ln{\left(\rho e^{i\theta}\right)}}{2\pi i}$. The product of these eigenvalues is proportional to 
	\begin{align}
	\prod_{n\ge 0}\left(1 - \frac{\left(\ln{(\rho e^{i\theta})}\right)^{2}}{\left(  \left(2n +1\right)\pi i\right)^{2}}\right) 
	\label{prods}
\end{align}
where we have normalised against the free theory partition function (without the insertion of $\tilde{\phi}\mathcal{A}\phi$)\footnote{Had we not made the field redefinition (\ref{redef}) then we would have instead considered the eigenvalues of $\left(\frac{d}{dt} -i \mathcal{A} + \frac{i\theta}{2\pi}\right)$. The homogeneous solution to this equation would be $\tilde{g}(\tau) = \mathscr{P}\exp{\left(\int_{0}^{\tau} i\mathcal{A}(\tau) + \frac{i\theta}{2\pi}\,d\tau\right)}$ which factorises to provide $\tilde{g}(2\pi) = e^{i\theta}g(2\pi)$. The eigenvalues of this operator are then $\tilde{\rho} = \rho e^{i\theta}$ and, imposing anti-periodic boundary conditions on the eigenfunctions, we would have arrived at (\ref{prods}).}. We have also paired the positive and negative integers with one another to produce the product over $n \geq 0$ which serves to regulate the determinant (we could just as well have used $\zeta$-function regularisation as in \cite{Hawk, Me2un, Me2Th} which would have led to the same expression).

The above expressions are familiar from the infinite product expansion of the $\cos$ function and it remains to finally take the product over $\rho$. We then arrive at an expression for the partition function written in terms of the determinant of quantities related to the Wilson-loop: 
	\begin{align}
	\mathcal{Z}\left[\mathcal{A}, \theta\right] \propto \det{ \left(\sqrt{e^{i\theta}g\left(2\pi\right)} + 1/\sqrt{e^{i\theta}g\left(2\pi\right)}\right)}  .
	\label{dets}
\end{align}
This is an explicit realisation of how the $\tilde{\phi}$, $\phi$ theory is related to the path ordered exponential we started with. In order to proceed we now show how this determinant can be written in terms of group invariant objects which turn out to be traces of the Wilson loop in different representations. 

Following \cite{Me1} we use a constant $SU(N)$ transformation to rotate $g(2\pi)$ onto the Cartan subalgebra so that
\begin{equation}
	g(2\pi) = \exp(\alpha_{i}H_{i});\qquad i = 1, \ldots, N - 1
\end{equation}
which allows us to write its eigenvalue equation in terms of the weights of the representation in which it is chosen to transform. The determinant in (\ref{dets}) can be expressed as the product over the eigenvalues of the matrix in the brackets which can also be expressed in terms of these weights. From the resulting expression we can then arrange the terms in to collections of group invariants constructed out of $g(2\pi)$. For example, the group $SU(3)$ was considered in \cite{Paul2} with the fields $\tilde{\phi}$ and $\phi$ taken to transform in the (conjugate) fundamental representation, $\mathbf{\bar{3}}$ and $\mathbf{3}$, without the presence of the $U(1)$ field. If we now include the factors of $\theta$ that arise in our current work then we find for the partition function
\begin{equation}
	\mathcal{Z}_{\mathbf{3}}	\left[\mathcal{A}, \theta\right] \propto e^{\frac{3}{2} i\theta} + \tr({g_{\mathbf{3}})} e^{\frac{1}{2}i\theta}+ \tr({g_{\mathbf{\bar{3}}})}e^{-\frac{1}{2}i\theta} + e^{-\frac{3}{2}i\theta},
	\label{SU3}
\end{equation}
where the subscript denotes the representation in which the trace is to be taken. Each term is interpreted as describing the interaction between the gauge field and a matter field which transforms in the labelled representation of the gauge group, whilst also carrying an exponent which denotes the $U(1)$ worldline hypercharge. We recognise the traces involved are of the super-Wilson loop in the representations constructed out of fully anti-symmetric tensor products of the fundamental -- in Young Tableaux notation these are \Yvcentermath1 {\Tiny$\bullet$}, $\Tiny{ \yng(1)}$, $ \Tiny{ \yng(1,1)}$ and $\Tiny{ \yng(1,1,1)}$  \Yvcentermath0.

\smallskip
Similarly, by taking $\tilde{\phi}$ and $\phi$ to transform in the fundamental of $SU(5)$ we have previously shown \cite{Me1} that the partition function evaluates to
\begin{equation}
	\mathcal{Z}_{\mathbf{5}}	\left[\mathcal{A}, \theta\right]\propto e^{\frac{5}{2} i \theta} + \mathrm{tr}\left(g_{\mathbf{5}}\right)e^{\frac{3}{2} i \theta} +  \mathrm{tr}\left(g_{\mathbf{10}}\right)e^ {\frac{1}{2}i \theta} + \mathrm{tr}\left(g_{\mathbf{\overbar{10}}}\right)e^ {-\frac{1}{2}i \theta} + \tr\left(g_{\mathbf{\bar{5}}}\right)e^{-\frac{3}{2} i \theta}  + e^{-\frac{5}{2} i \theta} 
	\label{SU5}
\end{equation}
which is easily seen to consist of traces of $g(2\pi)$ in the fully anti-symmetric representations\footnote{These can be represented as  \Yvcentermath1 {\Tiny$\bullet$}, $\Tiny{ \yng(1)}$, $ \Tiny{ \yng(1,1)}$, $\Tiny{ \yng(1,1,1)}$, $\Tiny{ \yng(1,1,1,1)}$ and $\Tiny{ \yng(1,1,1,1,1)}$ \Yvcentermath0.} of $SU(5)$. These two models are relevant to standard model physics and the unified theories based on $SU(5)$, flipped $SU(5)$ and $SO(10)$. In \cite{Me1} we have also described a way to provide chirality to the particle multiplets which appear and considered the more complicated case where the Grassmann fields transform in an arbitrary representation of the symmetry group.

For the remainder of this paper we will consider the general case that the symmetry group is $SU(N)$ and restrict our attention to the case that $\tilde{\phi}$ and $\phi$ transform in the conjugate fundamental and fundamental representations $\mathbf{\bar{N}}$ and $\mathbf{N}$. We also adopt a notation where the representations in which the traces of the Wilson loop are to be taken are indicated by Young Tableaux. This allows for general formulae to be presented which are valid for all choices of $N$ and avoids the clutter caused in a scheme based on indication of the dimensions of the representations involved. 

The generalisation of (\ref{SU3}) and (\ref{SU5}) is easy to work out by consideration of the Hilbert space structure discussed in section \ref{Fock}. We anticipate a sum over traces of the Wilson-loop taken in all completely anti-symmetric representations formed out of the fundamental representation, which are just the one column Young Tableaux of $SU(N)$. We write this as
\begin{equation}\Yvcentermath1
	\mathcal{Z}_{\mathbf{N}}	\left[\mathcal{A}, \theta\right]\propto \tr g( { \,\boldsymbol{\cdot}\,} ) e^{\frac{N}{2} i \theta} + \mathrm{tr}g({\,\Tiny\yng(1)\, } )e^{\frac{N-2}{2} i \theta} + \mathrm{tr}g({\,\Tiny\yng(1,1)\, } )e^ {\frac{N-4}{2}i \theta} + \ldots + \tr g( \underset{ \,\Tiny\yng(1,1)\, }{\overset{ \,\Tiny\yng(1,1)\, } {\rvdots} } )e^{-\frac{N-2}{2} i \theta} + \tr g({ \,\boldsymbol{\cdot} \,} ) e^{-\frac{N}{2} i \theta}\,,\Yvcentermath0
	\label{SUN}
\end{equation} 
where we have denoted the representation in which the trace is to be taken by its Young Tableau, so that the the term with $U(1)$ charge $\frac{N - 2p}{2}$ is denoted by the one-column Young Tableau with $p$ rows signifying that the matter field has $p$ completely anti-symmetric indices. To see why (\ref{SUN}) is correct we can consider the determinant (\ref{dets}). If $g(2\pi)$ transforms under the fundamental representation, $\mathbf{N}$, then it has $N$ eigenvalues, $\{\rho_{j}\}$. The partition function is then calculated by taking the product of these eigenvalues, and has the form
\begin{align}
	\mathcal{Z}_{\mathbf{N}}	\left[\mathcal{A}, \theta\right] &\propto  \prod_{j =1}^{N}\left(e^{\frac{1}{2}i\theta} \rho_{j}^{\frac{1}{2}}+ e^{-\frac{1}{2}i\theta} \rho_{j}^{-\frac{1}{2}}\right) \nonumber \\
	&\propto e^{-\frac{N}{2}i\theta}\prod_{j =1}^{N}\left(e^{i\theta} \rho_{j}^{\frac{1}{2}} + \rho_{j}^{-\frac{1}{2}}\right).
\end{align}
The coefficient of the term in the product with factor $e^{p i \theta}$ is determined by a sum involving the $\rho_{j}$ whose total number of terms is given by the binomial coefficient $d_{p} = {^{N}C_{p}}$, which we can use to deduce the composition of the contribution with hypercharge $\frac{N - 2p}{2}$. We know that it must be made up of group invariants constructed out of $g(2\pi)$ in totally anti-symmetric representations. However the dimension of the representation with $p$ totally anti-symmetrised indices is precisely $d_{p} = \frac{N!}{p!(N - p)!}$ and the trace of $g(2\pi)$ in this representation will consist of $d_{p}$ terms. This allows us to identify the coefficient of $e^{\frac{N - 2p}{2}i\theta}$ as the trace of $g(2\pi)$ taken in the representation with $p$ anti-symmetrised indices, as we have claimed in (\ref{SUN}) and in agreement with the analysis of section \ref{secStruct}. 

\subsection{Projecting onto representations}
Having completed the calculation of the restricted partition function we return to (\ref{Z}) to consider the remaining integrals. In particular we focus on the integral over the $U(1)$ modulus, $\theta$, whose purpose is to project intermediate states of the path integral over $\tilde{\phi}$ and $\phi$ onto fixed occupation number. We will now see how the integral over $\theta$ picks out just one of the representations of (\ref{SUN}) depending on the choice of the quantised Chern-Simons level $n - \frac{d_{R}}{2}$. We consider ($d_{R} = N$)
\begin{equation}
	\int_{0}^{2\pi} \frac{d\theta}{2\pi}\, e^{i(n -\frac{N}{2}) \theta } \mathcal{Z}_{\mathbf{N}}	\left[\mathcal{A}, \theta\right] 
\end{equation}
and for convenience we take out a factor of $e^{\frac{N}{2} i\theta}$ from (\ref{SUN}) so as to leave the simplified expression
\begin{align}\Yvcentermath1
	\int_{0}^{2\pi} \frac{d\theta}{2\pi}\, e^{i n\theta } \bigg(\tr g({ \,\boldsymbol{\cdot}  \,}) + \mathrm{tr} g( {\,\Tiny\yng(1)\, } )e^{-i \theta} + \mathrm{tr}g({\,\Tiny\yng(1,1)\, }) e^ {-2 i \theta} &+ \ldots + \tr g(\underset{ \,\Tiny\yng(1,1)\, }{\overset{ \,\Tiny\yng(1,1)\, } {\rvdots} })e^{-p i \theta}+\ldots  \nonumber\\
	&+\, \tr g(\underset{ \,\Tiny\yng(1,1)\, }{\overset{ \,\Tiny\yng(1,1)\, } {\rvdots} } )e^{-(N -1) i \theta} + \tr g({ \,\boldsymbol{\cdot}  \,})e^{-i N\theta}\bigg)\,,
	\label{project}
\end{align}
where the general term with exponent $\exp{(-p i \theta)}$ involves the trace of $g(2\pi)$ in the representation with $p$ totally anti-symmetric indices represented by a one column Young Tableau with $p$ rows. It is easy to see that enumerating through $n \in \{0 \ldots N\}$ the integral over $\theta$ picks out the representation in which the matter fields have $n$ anti-symmetric indices. In particular, setting $n = 1$ provides
\begin{equation}
	\tr (g_{\mathbf{N}}) = \tr_{\mathbf{N}} \mathscr{P}\exp{\left(i\int_{0}^{2\pi} \mathcal{A}[\omega(\tau), \psi(\tau)] d\tau\right)}
\end{equation}
which is the Wilson-loop coupling between the gauge field and a matter field transforming in the fundamental representation of $SU(N)$. It is of course sometimes necessary to include matter fields transforming in other representations of the symmetry group. One example is in unified theory; for $SU(5)$ unification the left-handed standard model particles are placed into the $\mathbf{\bar{5}}$ and $\mathbf{10}$, which have $4$ and $2$ totally antisymmetrised indices respectively. We have discussed this, and the inclusion of chirality, in \cite{Me1}.

Putting this path ordered exponential back into the full worldline partition function, (\ref{Z}), generates the well-known first quantised description of the partition function of a spinor field transforming in the fundamental representation of the symmetry group in the presence of the boson $A$ which gauges this symmetry:
\begin{equation}
	\int_{0}^{\infty} \frac{dT}{T} \oint \mathscr{D}\omega \mathscr{D}\psi\, e^{-\frac{1}{2} \int_{0}^{2\pi} \frac{\dot{\omega}^{2}} {T} + \psi \cdot \dot{\psi} \, d\tau} \,\tr_{\mathbf{N}}  \mathscr{P}\exp{\left(i\int_{0}^{2\pi} \mathcal{A}[\omega(\tau), \psi(\tau)] d\tau\right)}.
\end{equation}
At this point it is important to reiterate that the procedure outlined above has been considered in perturbative worldline calculations before \cite{Bastwl1, Bastwl2}. Furthermore, as we have previously demonstrated (see \cite{Paul2, Me1}) it is sometimes advantageous not to carry out any projection at all, instead including the contribution from all anti-symmetric representations. We believe, however, that this is the first time that the complete path integral has been computed analytically in this context. A natural question to ask at this point is whether or not the method can be adapted or extended in order to generate the Wilson-loop coupling for a matter field that transforms in representations other than those constructed as totally anti-symmetric tensor products whilst still only using color fields transforming in the (anti-)fundamental representation of the chosen symmetry group. In the following sections we explain how to generate couplings for fields transforming in arbitrary representations of the symmetry group. This question is important for first quantised descriptions of such matter fields where methods have been, up to now, somewhat limited, being restricted to representations with special symmetries.

\section{Mixed Symmetry tensors}
\label{secMixed}
In this section we consider how to describe matter fields that transform in an arbitrary representation of the symmetry group in the presence of the background gauge field. To do so we will use generalise the approach taken above by using tensor products of anti-symmetric representations to build a Young Tableau with a chosen shape. This will require the use of further copies of the additional Grassmann fields transforming in the same representatations as above. Using such ideas a worldline description of higher spin fields has been described before in a phase space formulation \cite{Bastvt, hspin, hspin2}. A similar method has been used in the context of differential forms on complex manifolds \cite{Bastforms, Bastforms2, Bastforms3} and to construct detour complexes from BRST quantisation of worldlines theories \cite{Bastbrst}. We shall follow the general approach taken by these authors but it must be stressed that the focus of this article is to reproduce the correct interaction between the matter fields and the background gauge field. We also continue to work in configuration space and will reproduce path ordered exponentiated line integrals to realise the Wilson-loop coupling which describes the interaction. We leave a perturbative calculation of scattering amplitudes for future work.

The result of the previous section was a worldline theory that produces a sum over all fully anti-symmetric representations of a chosen symmetry group (strictly speaking these are representations built out of fully anti-symmetric tensor products of the representation in which the Grassmann fields $\tilde{\phi_{r}}$ and $\phi_{r}$ transform \cite{Me1}). To combine multiple anti-symmetric representations requires the introduction of separate families of the anti-commuting fields -- we showed this in \cite{JOdof}. Each family will generate its own set of interactions between the matter field and the background gauge field which can then be combined to generate more complicated representations by forming their tensor-product. To see how this is incorporated in the worldline theory we return to (\ref{worldlinephi}). As discussed above this consists of a factor describing the free dynamics of the matter field and a second piece responsible for producing the interaction with the background field. It is this second part that will be modified in the next section.

\subsection{The generalised worldline theory}
The adjustment we proposed in \cite{JOdof} is to introduce $F$ families of the Grassmann fields $\tilde{\phi}_{k}^{r}$ and $\phi_{kr}$, denoting each family with an index $k \in \{1, \ldots, F\}$. The generalisation of (\ref{worldlinephi}) is then
\begin{equation}
	 S[\omega, \psi, \tilde{\phi}, \phi] =\int_{0}^{2\pi} d\tau \left[ \frac{\dot{\omega}^{2}}{2T} + \frac{i}{2}\psi \cdot \dot{\psi} + i\tilde{\phi}^{r}_{k} \dot{\phi}_{kr} + \tilde{\phi}_{k}^{r} \mathcal{A}_{r}{}^{s} \phi_{ks}\right].
	 \label{worldlinephiF}
\end{equation}
which consists of $F$ copies of the interaction between the matter field and the gauge field. The Fock space of this extended theory is much richer than before, since creation operators associated with different families can act independently to populate the physical state space. The action (\ref{worldlinephiF}) can incorporate the super-symmetry discussed in section \ref{secStruct} ($F$ families of super-partners $\tilde{z}_{k}^{r}$ and $z_{kr}$ are needed) but the invariance under unitary transformations of the anti-commuting fields is enriched. It is now possible to make a global $U(F)$ transformation on the fields $\tilde{\phi}_{k}$ and $\phi_{k}$ which rotates between the families. This is generated by the constant matrix $\alpha_{kl}$ and takes the following infinitesimal form
\begin{equation}
	\delta_{\alpha}\tilde{\phi}_{k}^{r} =  i \tilde{\phi}^{r}_{l}\alpha_{lk}; \qquad \delta_{\alpha}\phi_{kr} =  -i \alpha_{kl}\phi_{lr}.
\end{equation}
The global symmetry implies the existence of the conserved currents $N_{kj} = \tilde{\phi}_{k}^{r} \phi_{jr}$ which upon quantisation become the generalisation of the occupation number operator in the previous section. The diagonal elements $N_{kk}$ just give the occupation number of the $k$th family and generate independent $U(1)$ transformations, whilst the off-diagonal elements step between families.

This new symmetry can be gauged via the introduction of fields $a_{kj}(\tau)$ so as to form a covariant derivative $D_{kj} = (\delta_{kj}\frac{d}{d\tau} + ia_{kj})$, from which the following worldline action can be constructed
\begin{equation}
	 S[\omega, \psi, \tilde{\phi}, \phi] = \int_{0}^{2\pi} d\tau \left[ \frac{\dot{\omega}^{2}}{2T} + \frac{i}{2}\psi \cdot \dot{\psi} + \tilde{\phi}_{k}^{r}(i\delta^{s}_{r}D_{kj} + \mathcal{A}_{r}{}^{s}\delta_{kj}) \phi_{js} \right],
\end{equation}
which is invariant under the full non-Abelian symmetry group if $\delta_{\alpha}a_{kj} = \dot{\alpha}_{kj} - i[\alpha, a]_{kj}$ transforms in the adjoint representation of $U(F)$. However, for reasons which will now be explained we found in \cite{JOdof} it is not advantageous to gauge the full symmetry group and it will prove necessary to consider only a partial gauging, leaving part of the subgroup $U(1)^{F} \subset U(F)$ invariant.

In the previous section the gauging of the group allowed for the introduction of a Chern-Simons term for $a(\tau)$ which played the r\^{o}le of projecting the intermediate states of the path integral onto fixed occupation number. This ensured that from the list of representations generated by the Grassmann fields only one wavefunction component contributed to the final result. Now that there are multiple families of these anti-commuting fields this projection is more complex, since a mechanism is needed to fix the number of indices contributed to the matter field by each family separately. The problem with gauging the whole $U(F)$ group is that the only gauge invariant quantity that can be constructed is $\tr{(a)} = \sum_{k = 1}^{F} a_{kk}$ which allows for the introduction of only one Chern-Simons term  \cite{Bastbrst}
\begin{equation}
	S[a] = \int d\tau \,s\, \tr{(a(\tau))}.
\end{equation}
This fixes the occupation numbers of each family to be the same and imposes no constraints on how the families are combined. This does not allow us to specify the indices in each family independently and will not lead to a projection onto an irreducible representation. 

The first problem can be overcome by instead fixing only the Abelian subgroup $U(1)^{F}$. This ensures that the diagonal elements of $a_{kj}$ transform as a total derivative so we can construct independent Chern-Simons terms for each family \cite{Bastbrst}
\begin{equation}
	S[a] = \int d\tau \, \sum_{k=1}^{F}s_{k}\, a_{kk}\left(\tau\right).
\end{equation}
This can be used to select the occupation numbers of each family separately. However, this still does not achieve irreducibility because there remains too much freedom in how these representations can be combined. We presented the resolution of this problem in \cite{JOdof} where we found that one should in fact gauge only those $N_{kj}$ with $k \geqslant j$. This retains the gauge invariance of the independent Chern-Simons terms above since in this instance $\delta a_{kk} = \dot{\alpha}_{kk
}\,\, \forall k$. We showed that the equations of motion for the fields $a_{kj}$, in combination with the above Chern-Simons terms imply the vanishing off the off-diagonal number operators $N_{kj}$ which constrains how the creation operators of each family can be combined to build up the physical state space. This will later be seen to reflect the Lie algebra rules for combining tensor products of representations in such a way as to arrive at a projection onto an irreducible representation. We are consequently led to the following action for $F$ families of colour fields which represents the gauging of the ``auxiliary group'' generated by the upper triangular elements $\alpha_{k\geqslant j}$
\begin{equation}
	S[\tilde{\phi}, \phi, a] = \int_{0}^{2\pi} d\tau \left[ i\tilde{\phi}^{r}_{k} \dot{\phi}_{ks} +\tilde{\phi}^{r}_{k} \mathcal{A}_{r}{}^{s}\phi_{kr} -\textstyle{\sum\limits_{k=1}^{F}}a_{k}\left(N_{k} - s_k\right)- \textstyle{\sum\limits_{j < k}}a_{kj}N_{kj}\right],
\end{equation}
where, as in \cite{JOdof}, we have introduced the notation $a_{k} \equiv a_{kk}$ and we have separated the diagonal generators of the auxiliary gauge group from the off-diagonal entries so as to pair them with the Chern-Simons terms. A simple calculation of the equations of motion for the $a_{k}$ and $a_{kj}$ is sufficient to show that these fields impose the constraints
\begin{equation}
	\left(\tilde{\phi}^{r}_{k}\partial_{\tilde{\phi}^{r}_{k}} -n_{k}\right)\Psi(x, \tilde{\phi}) = 0 \qquad \textrm{and} \qquad \tilde{\phi}^{r}_{k}\partial_{\tilde{\phi}^{r}_{j}}\Psi(x, \tilde{\phi}) = 0
\end{equation}
for $k$, $j \in \{1, 2, \ldots, F\}$ (there is no sum over these indices). These are the conditions that set the occupation number of each family of colour fields and achieve irreducibility respectively. This ensures that the remaining wavefunction component has the correct symmetries to transform in a single irreducible representation. For the rest of this section we will focus on this action and its functional quantisation, making use of the results found above.

\subsection{Gauge fixing and functional quantisation}
Before carrying out the functional integration over the $F$ families of colour fields we must gauge fix to take into consideration the overcounting caused by the symmetry of the action under the auxiliary gauge group. As in the $U(1)$ case previously discussed the auxiliary $U(F)$ symmetry can at best be fixed by setting $a_{kj}$ to be constant (see Appendix~\ref{app:B}) which, as explained in \cite{JOdof}, can be taken to be in the form
\begin{equation}
	2\pi \hat{a}_{kj} = \begin{pmatrix}
		\theta_{1} & 0 & \cdot & 0 \\ 0 & \theta_{2} & \cdot & 0 \\ \cdot & \cdot & \cdot & \cdot \\ 0 & 0 & \cdot & \theta_{F}
	\end{pmatrix}
	\label{afix}
\end{equation}
where the $\{\theta_{k}\}$ are angular moduli to be integrated over. This gauge fixing can be compensated for by the introduction of the Faddeev-Popov measure which we denote by $\mu\left(\{\theta_{k}\}\right)$. This is easily determined from the infinitesimal transformation of $a$:
\begin{equation}
	\mu\left(\{\theta_{k}\}\right) = \left.\Det{\left(\frac{d}{d \tau} + i [a, \cdot]\right)}\right|_{a = \hat{a}}
	\label{FP}
\end{equation}
and depends on how the global $U(F)$ symmetry is gauged. For the partial gauging used in \cite{JOdof} which we discussed above we found the modular measure
\begin{equation}
	\mu\left(\{\theta_{k}\}\right) = \prod_{j<k}\mu\left(\{\theta_{k}, \theta_{j}\}\right)= \prod_{j< k}2i \sin{\left(\frac{\theta_{j} - \theta_{k}}{2}\right)}
	\label{mufp}
\end{equation}
which we will show is responsible for the required projection.


In what follows we again focus on the $\tilde{\phi}$ and $\phi$ theory to construct a worldline representation of matter fields with arbitrary representation coupled to a background gauge field. For path integral quantisation we again rotate to Euclidean space. On the gauge slice defined by $a_{kj} = \hat{a}_{kj}$, the generalisation of the field theory partition function (\ref{Z}) is now
\begin{align}
	\int_{0}^{\infty} \frac{dT}{T} \oint \mathscr{D}\omega \mathscr{D}\psi\, e^{-\frac{1}{2} \int_{0}^{2\pi} \frac{\dot{\omega}^{2}} {T} + \psi \cdot \dot{\psi} \, d\tau }K_{F}\prod_{k = 1}^{F}\int_{0}^{2\pi} \frac{d\theta_{k}}{2\pi}\, e^{is_{k} \theta_{k} } \mu\left(\{\theta_{k}\}\right) \mathcal{Z}^{(F)}\left[\mathcal{A},\{\theta_{k}\}\right]
	\label{ZF}
\end{align}
where the reduced partition function (which produces the Wilson-loop interaction between the matter fields and the gauge fields) on the chosen gauge slice is
\begin{equation}
	\mathcal{Z}^{(F)}\left[\mathcal{A}, \{\theta_{k}\}\right] = \prod_{k=1}^{F} \int\mathscr{D}\tilde{\phi}_{k} \mathscr{D}\phi_{k} \, e^{-\int_{0}^{2\pi} d\tau\, \tilde{\phi}_{k}^{r}(\delta^{s}_{r}D_{kj} -i \mathcal{A}_{r}{}^{s}\delta_{kj}) \phi_{js}},
	\label{ZaF}
\end{equation}
which will be the focus of the remainder of this section. In the first equation above, $K_{F}$ is a normalising constant equal to the inverse of the number of fundamental domains \cite{JOdof} and the $s_{k}$ are the Chern-Simons levels fixing the occupation number of each family of colour fields. Noting that with (\ref{afix}) the covariant derivative is diagonal the functional integration factorises and we can repeat the arguments of section \ref{secFunc} over each family. The boundary conditions on each family of fields remain anti-periodic. The only additional fact is that each family carries its own $U(F)$ modulus $\theta_{k}$ which, if desired, can be absorbed into the Grassmann fields $\tilde{\phi}_{k}$ and $\phi_{k}$ by making field redefinitions akin to (\ref{redef}). The functional determinant which arises upon integrating over $\{\tilde{\phi}_{k}\}$ and $\{\phi_{k}\}$ can then be written as a product of determinants related to the super-Wilson loop which generalises (\ref{dets}):
\begin{align}
	\mathcal{Z}^{(F)}\left[\mathcal{A}, \{\theta_{k}\}\right] \propto \prod_{k=1}^{F} \det{ \left(\sqrt{e^{i\theta_{k}}g\left(2\pi\right)} + 1/\sqrt{e^{i\theta_{k}}g\left(2\pi\right)}\right)}.
\label{detsF}
\end{align}
Furthermore, each factor in this product has been calculated above, leading simply to a sum of traces of the super-Wilson loop over all representations constructed out of fully anti-symmetric tensor products with a hypercharge that is related to the number of indices associated to the representation -- that is, products of the form (\ref{SUN}). It is now time to return to the integral over the $U(F)$ moduli which is needed to enforce irreducibility.

\subsection{Irreducibility}
The functional integration over each family of Grassmann variables results in a reducible sum of products of traces of the super-Wilson loop in anti-symmetric representations. In this section we understand the effect of the measure $\mu\left(\{\theta_{k}\}\right)$ to see how the partial gauging of the $U(F)$ symmetry fixes the representation contributed by each family and the group structure of the product that emerges. For simplicity we first focus on the case of two families, $F = 2$, which will illustrate the important behaviour, before generalising the results to an arbitrary choice of $F$. 

The Chern-Simons terms in (\ref{ZF}) allow the specification of the number of anti-symmetric indices (or number of rows in the one-column Young Tableaux) in each family, denoted by $n_{k}$. We will denote our choices of the $n_{k}$ by n-tuples $n = (n_{1}, n_{2}, \ldots, n_{F})$ where without loss of generality we choose $n_{k+1} \geqslant n_{k}$. For $F = 2$ and a symmetry group $SU(N)$ we take $0 \leqslant n_{1} \leqslant n_{2} \leqslant N$. Following \cite{JOdof} this requires us to take the Chern-Simons levels to be $s_{1} = n_{1} - \frac{N}{2} -\frac{1}{2}$ and $s_{2} = n_{2} - \frac{N}{2} + \frac{1}{2}$ which leads us to consider ($K_{2} = 1$)
\begin{equation}
	\int_{0}^{2\pi} \frac{d\theta_{1}}{2\pi}\int_{0}^{2\pi} \frac{d\theta_{2}}{2\pi}\, e^{i(n_{1} -\frac{N}{2}-\frac{1}{2}) \theta_{1} }e^{i(n_{2} -\frac{N}{2}+\frac{1}{2}) \theta_{2} }\mu\left(\{\theta_{1}, \theta_{2}\}\right) \mathcal{Z}^{(2)}_{\mathbf{N}}	\left[\mathcal{A}, \{\theta_{1}, \theta_{2}\}\right]. 
\end{equation}
For $\mathcal{Z}^{(2)}_{\mathbf{N}}$ we use two copies of (\ref{SUN}) with associated $U(1)$ hypercharges and following the steps that led to (\ref{project}) it is again useful to extract factors of $e^{\frac{N}{2}i \theta_{1}}$ and $e^{\frac{N}{2}i \theta_{2}}$ to cancel the same terms in the Chern-Simons moduli, leaving the integral\Yvcentermath1
\begin{align}
	\int_{0}^{2\pi} \frac{d\theta_{1}}{2\pi}&\int_{0}^{2\pi} \frac{d\theta_{2}}{2\pi}\,  e^{i \left(n_{1}-\frac{1}{2}\right) \theta_{1} }e^{i \left(n_{2}+\frac{1}{2}\right) \theta_{2} }\left(e^{i\frac{\theta_{1}}{2}}e^{-i\frac{\theta_{2}}{2}} - e^{-i\frac{\theta_{1}}{2}}e^{i\frac{\theta_{2}}{2}}\right) \times \nonumber \\
	 &\bigg(\tr g({ \,\boldsymbol{\cdot}  \,}) + \mathrm{tr} g( {\,\Tiny\yng(1)\, } )e^{-i \theta_{1}} + \mathrm{tr}g({\,\Tiny\yng(1,1)\, }) e^ {-2 i \theta_{1}} + \ldots  + \tr g(\underset{ \,\Tiny\yng(1,1)\, }{\overset{ \,\Tiny\yng(1,1)\, } {\rvdots} } )e^{-(N -1) i \theta_{1}} + \tr g({ \,\boldsymbol{\cdot}  \,})e^{-i N\theta_{1}}\bigg) \times \nonumber \\
	 &\bigg(\tr g({ \,\boldsymbol{\cdot}  \,}) + \mathrm{tr} g( {\,\Tiny\yng(1)\, } )e^{-i \theta_{2}} +\mathrm{tr}g({\,\Tiny\yng(1,1)\, }) e^ {-2 i \theta_{2}} + \ldots  + \tr g(\underset{ \,\Tiny\yng(1,1)\, }{\overset{ \,\Tiny\yng(1,1)\, } {\rvdots} } )e^{-(N -1) i \theta_{2}} + \tr g({ \,\boldsymbol{\cdot}  \,})e^{-i N\theta_{2}}\bigg), \Yvcentermath0
	 \label{projectF}
	\end{align}
which we use to explore the behaviour of the worldline theory in this simple case.

Integrating over the $U(2)$ moduli produces a sum of products of the trace of the Wilson loop in different representations. The traces combine to form the desired result, being the trace of the Wilson loop in a single, chosen irreducible representation. For example it is easy to check that, taking $n = (1, 1)$, the integration provides the correct combination of Wilson loop interactions
\begin{equation}\Yvcentermath1
	\tr g({\,\Tiny\yng(1)\, }) \times \tr g({\,\Tiny\yng(1)\, })-\tr g({\,\Tiny\yng(1,1)\, })\times \tr g({ \,\boldsymbol{\cdot}  \,})  = \tr g({\,\Tiny\yng(2)\, }) 
	\Yvcentermath0
	\label{eg2'}
\end{equation}
where the first trace in each product arises from integrating over $\theta_{2}$ and the second from integrating over $\theta_{1}$.  Similarly for $n = (2, 3)$ we get
\begin{equation}\Yvcentermath1
	\tr g({\,\Tiny\yng(1,1,1)\, }) \times \tr g({\,\Tiny\yng(1,1)\, }) - \tr g({\,\Tiny\yng(1,1,1,1)\, }) \times \tr g({\,\Tiny\yng(1)\, }) = \tr g({\,\Tiny\yng(2,2,1)\, }) 
	\Yvcentermath0
	\label{eg1'}
\end{equation}
which can be verified to correctly generate the trace of the Wilson-loop in the representations whose Young-Tableau has $n_{1}$ rows in the first column and $n_{2}$ rows in the second column\footnote{To demonstrate that the measure plays a non-trivial r\^{o}le we note that if we set $\mu\left(\{\theta_{1}, \theta_{2}\}\right) = 1$, corresponding to the Faddeev-Popov determinant associated with the gauging of the $U(1)^{F}$ Abelian subgroup of the full $U(F)$ symmetry group, then the above two results become
\begin{equation}\Yvcentermath1
	\tr g({\,\Tiny\yng(1)\, }) \times \tr g({\,\Tiny\yng(1)\, }) = \tr g({\,\Tiny\yng(2)\, }) + \tr g({\,\Tiny\yng(1,1)\, }),
	\Yvcentermath0
	\label{eg2}
\end{equation}
consisting of the trace in both the fully symmetric and fully anti-symmetric representations, and
	\begin{equation}\Yvcentermath1
	\tr g({\,\Tiny\yng(1,1,1)\, }) \times \tr g({\,\Tiny\yng(1,1)\, }) = \tr g({\,\Tiny\yng(2,2,1)\, }) + \tr g({\,\Tiny\yng(2,1,1,1)\, }) + \tr g({\,\Tiny\yng(1,1,1,1,1)\, }),
	\Yvcentermath0
	\label{eg1}
\end{equation}
neither of which has achieved the desired irreducibility. Clearly some extra structure is needed in order to constrain the result to pick out a single, irreducible representation from the product -- this is the job of the \textit{partial} gauging of the unitary symmetry and the associated Faddeev-Popov measure. These equalities, and those in the main text, are easily verified by again rotating $g(2\pi)$ onto the Cartan subalgebra and finding the sum of its eigenvalues in each representation.}. So with this approach it is now possible to project onto any irreducible representation of the gauge group which can be denoted by two-column Young Tableaux simply be specifying the number of rows in each column and making use of the measure on $U(F)$ moduli in (\ref{mufp}). 

\subsection{Projection onto an arbitrary representation}
The generalisation of these results to an arbitrary number of families is fairly straightforward. To project onto a representation denoted by an $F$-column Young Tableaux we specify the number of rows, $n_{k}$, in each column and include the Chern-Simons modular term $\exp{(i s_{k} \theta_{k})}$ for each family. The generalisation of the $U(F)$ measure $\mu\left(\{\theta_{k}\}\right)$ is given in (\ref{mufp}), although it could also be arrived at by symmetry considerations as follows. Each family should be treated equally and no particular pairing of the families should be favoured so we need a copy of $\mu\left(\{\theta_{k}, \theta_{j}\}\right)$ for each possible pairing. Indeed, doing so leads to
\begin{equation}
	\mu\left(\{\theta_{k}\}\right) = \prod_{j<k}\mu\left(\{\theta_{k}, \theta_{j}\}\right)=\prod_{j< k}2i \sin{\left(\frac{\theta_{j} - \theta_{k}}{2}\right)}
\end{equation}
which takes the product of the $U(2)$ measure across all pairs and correctly coincides with the Faddeev-Popov determinant for the partial gauging of the $U(F)$ symmetry as presented above. So in the general case -- with each family of Grassmann fields taken to transform in the fundamental representation of $SU(N)$ -- (\ref{ZF}) becomes
\begin{align}\Yvcentermath1
	&\int_{0}^{\infty} \!\frac{dT}{T} \!\oint \!\mathscr{D}\omega \mathscr{D}\psi\, e^{-\frac{1}{2} \int_{0}^{2\pi} \!\frac{\dot{\omega}^{2}} {T} + \psi \cdot \dot{\psi} \, d\tau}K_{F}\prod_{k = 1}^{F}\int_{0}^{2\pi} \frac{d\theta_{k}}{2\pi}\, e^{is_{k} \theta_{k} } \prod_{j<k}\mu\left(\{\theta_{j}, \theta_{k}\}\right) \mathcal{Z}_{\mathbf{N}}^{(F)}\left[\mathcal{A},\{\theta_{k}\}\right] \,. \nonumber \\
\end{align}
This formula projects onto the correct representation by forming the tensor product of the representations produced by each family and then iteratively compensating for the unwanted representations that appear in their tensor product decomposition. That is, choosing occupation numbers $n = \left(n_{1}, n_{2}, \ldots, n_{F}\right)$, it forms the tensor product of representations whose Young Tableaux have one column and $n_{k}$ rows and then pairwise subtracts the representations whose Young-Tableaux do not consist of $n_{1} \leq n_{2} \leq \ldots \leq n_{F}$ rows in each column. As we have shown in \cite{JOdof}, this is realised by taking Chern-Simons levels $s_{k} = n_{k} - \frac{N}{2} - \frac{F-(2k-1)}{2}$, which allows the above formula to be written in the simpler form
\begin{align}
	&\int_{0}^{\infty} \!\frac{dT}{T} \!\oint \!\mathscr{D}\omega \mathscr{D}\psi\, e^{-\frac{1}{2} \int_{0}^{2\pi} \frac{\dot{\omega}^{2}} {T} + \psi \cdot \dot{\psi} \, d\tau}\prod_{k = 1}^{F}\int_{0}^{2\pi} \!\frac{d\theta_{k}}{2\pi}\, e^{i n_{k} \theta_{k} } \prod_{j<k}\left(1\! -\! e^{- i \theta_{j}}e^{i \theta_{k}}\right) \times \nonumber \\
	&\qquad \prod_{k = 1}^{F} \bigg(\tr g({ \,\boldsymbol{\cdot}  \,}) + \mathrm{tr} g( {\,\Tiny\yng(1)\, } )e^{-i \theta_{k}} + \mathrm{tr}g({\Yvcentermath1 \,\Tiny\yng(1,1)\, }) e^ {-2 i \theta_{k}} + \ldots  + \tr g(\underset{ \,\Tiny\yng(1,1)\, }{\overset{ \,\Tiny\yng(1,1)\, } {\rvdots} } )e^{-(N -1) i \theta_{k}} + \tr g({ \,\boldsymbol{\cdot}  \,})e^{-i N\theta_{k}}\bigg). 
	\Yvcentermath0
	\label{ZF'}
\end{align}
This is the first of the two main results of this article. To bring this formula into line with the form of the results presented in \cite{JOdof} it is necessary to make a change of variables by introducing $F$ complex parameters $z_{k} = e^{i \theta_{k}}$. These variables are in fact (worldline) Wilson-loops which parameterise inequivalent fields with respect to the auxiliary gauge group. Doing so allows (\ref{ZF'}) to be recast in a concise form as
\begin{align}
		\int_{0}^{\infty} \!\frac{dT}{T} \!\oint \!\mathscr{D}\omega \mathscr{D}\psi\, e^{-\frac{1}{2}\int \frac{\dot{\omega}^{2}} {T} + \psi \cdot \dot{\psi} \,d\tau}\prod_{k = 1}^{F}\oint \!\frac{dz_{k}}{2\pi i} \prod_{l<j}\left(1 - \frac{z_{j}}{z_{l}}\right)  \prod_{k= 1}^{F} \sum_{p_{k} = 1}^{N} \frac{\tr g(2\pi)_{[p_{k}]}}{z_{k}^{p_{k}+1-n_{k}}}
	\Yvcentermath0
	\label{ZF'z}
\end{align}
where the integrals with respect to $z_{k}$ are now over closed contours in the complex plane encircling the origin. These integrals simply pick out the poles that arise at $z = 0$ for certain values of the summation variables $p_{k}$ which denote the representation in which the traces of the super-Wilson loop are to be taken: by $\tr g(2\pi)_{[p_{k}]}$ we mean the trace of the super-Wilson loop in the representation with $p_{k}$ fully antisymmetric indices.

To illustrate the usage of (\ref{ZF'}) or (\ref{ZF'z}) we present the result for $F = 6$ families of Grassmann fields and choose $n = (1,2,4,4,6,7)$ (picking $N \ge 7$). Then by integrating over the $U(6)$ moduli we arrive at the following worldline path integral
\begin{equation}
	\int_{0}^{\infty} \frac{dT}{T} \oint \mathscr{D}\omega \mathscr{D}\psi\, e^{-\frac{1}{2} \int_{0}^{2\pi} \frac{\dot{\omega}^{2}} {T} + \psi \cdot \dot{\psi} \, d\tau} \,\tr_{R} \mathscr{P} \exp{\left(i\int_{0}^{2\pi} \mathcal{A}[\omega(\tau), \psi(\tau)] d\tau\right)}
\end{equation}
where the representation, $R$, in which the super-Wilson loop is taken to transform has Young Tableau
\begin{equation}\Yvcentermath1
	{\,\tiny \yng(6,5,4,4,2,2,1)\, }
	\Yvcentermath0
\end{equation}
in agreement with the specification of rows given by $n$. Of more interest is perhaps a projection onto the adjoint representation of the chosen gauge group. This requires the use of $F = 2$ families and the choice $n = (1, N-1)$ to project onto the irreducible representation with Young Tableau
\begin{equation}\Yvcentermath1
	{\underset{ \,\tiny\yng(1,1)\, }{\overset{\hphantom{\tiny\yng(1)} \,\tiny\yng(2,1)\, } {\rvdots}}}\, ,
	\Yvcentermath0
\end{equation}
where the first column has $N-1$ boxes so the dimension of the representation is $N^{2} - 1$. This is an important step towards developing a worldline description of gluons and their interaction with a background field.

The worldline theory (\ref{ZF'}) can be tested against any representation and it is instructive to do so for small values of $F$ to see explicitly how the modular measure enforces irreducibility by compensating for the proliferation of unwanted representations. For example, taking $F = 3$ and $n = ( 1, 1, 2)$ formula (\ref{ZF'}) gives
\begin{align}\Yvcentermath1
	\tr g( {\,\Tiny\yng(3,1)\, } ) = \,\,&\tr g( {\Yvcentermath1\,\Tiny\yng(1,1)\, } )\, \tr g( {\Yvcentermath1\,\Tiny\yng(1)\, } )\,\tr g( {\Yvcentermath1\,\Tiny\yng(1)\, } ) - \tr g( {\Yvcentermath1\,\Tiny\yng(1,1,1)\, } )\,\tr g( {\Yvcentermath1\,\Tiny\yng(1)\, } )\,\tr g( {\,\boldsymbol{\cdot}\, } ) \nonumber \\
	- \,&\tr g( {\Yvcentermath1\,\Tiny\yng(1,1)\, } )\, \tr g( {\Yvcentermath1\,\Tiny\yng(1,1)\, } )\, \tr g( {\,\boldsymbol{\cdot}\, } )\hskip 2.5px + \tr g( {\Yvcentermath1\,\Tiny\yng(1,1,1,1)\, } )\,\tr g( {\,\boldsymbol{\cdot}\, } ) \, \tr g( {\,\boldsymbol{\cdot}\, } ) 
	\Yvcentermath0
	\label{egF}
\end{align}
which can be analysed as follows. The first term on the right hand side generates a sum over irreducible representations 
\begin{align}\Yvcentermath1
	&\tr g( {\Yvcentermath1\,\Tiny\yng(1,1)\, } )\, \tr g( {\Yvcentermath1\,\Tiny\yng(1)\, } )\,\tr g( {\Yvcentermath1\,\Tiny\yng(1)\, } )= \tr g( {\Yvcentermath1\,\Tiny\yng(3,1)\, } ) + \tr g( {\Yvcentermath1\,\Tiny\yng(2,2)\, } ) + 2\, \tr g( {\Yvcentermath1\,\Tiny\yng(2,1,1)\, } ) + \tr g( {\Yvcentermath1\,\Tiny\yng(1,1,1,1)\, } )
	\Yvcentermath0
	\label{dec1}
\end{align}
whose trailing three terms are not desired. From this sum, the last term and one copy of the penultimate term are removed by the second product on the right hand side of (\ref{egF}):
\begin{align}\Yvcentermath1
	&\tr g( {\Yvcentermath1\,\Tiny\yng(1,1,1)\, } )\, \tr g( {\Yvcentermath1 \,\Tiny\yng(1)\, } ) = \tr g( {\Yvcentermath1\,\Tiny\yng(2,1,1)\, } ) + \tr g( {\Yvcentermath1\,\Tiny\yng(1,1,1,1)\, } ). 
	\Yvcentermath0
	\label{dec2}
\end{align}
The term with square tableaux in (\ref{dec1}) and the second copy of its penultimate term are countered by the first term on the second line of (\ref{egF}) but this also subtracts an extra copy of the representation with the one-column tableau
\begin{align}\Yvcentermath1
	&\tr g( {\Yvcentermath1\,\Tiny\yng(1,1)\, } )\, \tr g( {\Yvcentermath1 \,\Tiny\yng(1,1)\, } ) = \tr g( {\Yvcentermath1\,\Tiny\yng(2,2)\, } ) + \tr g( {\Yvcentermath1\,\Tiny\yng(2,1,1)\, } ) + \tr g( {\Yvcentermath1\,\Tiny\yng(1,1,1,1)\, } ). 
	\Yvcentermath0
	\label{dec3}
\end{align}
This last contribution must then be added on again, which explains the final term on the second line of (\ref{egF}). This procedure removes all the unwanted representations of (\ref{dec1}), leaving only the first term on the right hand side of that decomposition, thereby completing the projection onto the desired representation. 

Formula (\ref{ZF'}) allows complete freedom in the specification of the representation in which the matter field is to be taken in the first quantised representation of the quantum field theory. It overcomes the limitations of previous work and gives a general framework in which quantum mechanics can be used to describe a gauged field theory based on given Lie group. 

\section{Symmetric representations}
\label{secSymm}
In the above construction we took the auxiliary worldline fields, which generate the coupling of the matter field to the gauge field, to be anti-commuting. As such they spanned a Hilbert space whose wavefunction components transform in tensor products of fully anti-symmetric representations of the gauge group. In \cite{JOdof}, building upon \cite{Bastwl2}, we showed that a similar idea works when the colour fields are chosen to have bosonic statistics. In this case, the correct coupling to the gauge field will be built out of products of traces of the Wilson-loop interaction in fully \textit{symmetric} representations. The aim is to present a worldline theory which leads to a formula analogous to (\ref{ZF'}) which may be used in other first quantised settings. To do so, we follow our earlier work in \cite{JOdof}, which we briefly recap below.

Before doing so, we wish to point out that our earlier work in \cite{JOdof} suggests an immediate method of arriving at a worldline description of matter fields transforming in fully symmetric representations of the symmetry group interacting with the gauge field. A representation with $p$ fully symmetric indices has a Young Tableau with one row and $p$ columns. We could build this up by following the earlier sections with the use of $F = p$ families of Grassmann fields, using the worldline theory based on fully anti-symmetric representations of the gauge group. Simply by setting $n = (1, 1, \ldots, 1)$, the main formula (\ref{ZF'}) gives 
\begin{equation}\Yvcentermath1
	\tr g( { \,\Tiny\yng(2)}\!\!  \rhdots \!\! {\Tiny\yng(2)\, } )\Yvcentermath0,
\end{equation}
where the trace of the Wilson-loop is taken in the representation whose Young Tableau has one row with $p$ columns as desired. 

In the following sections, however, we consider an alternative approach to worldline descriptions of matter fields where the building blocks themselves are instead fully symmetric representations of the symmetry group. To effect this change it suffices to make just a small modification to the worldline theory. Rather than taking the colour fields to be Grassmann variables we instead follow \cite{Bastwl2} by taking these fields to be bosonic. We begin with a single family of the auxiliary fields as the result will be needed in the generalisation to multiples copies of the bosonic colour variables. So we consider the gauge group $SU(N)$ and define $N$ pairs of fields $\tilde{\phi}^{r}$ and $\phi_{r}$. We again impose the Poisson brackets $\{\tilde{\phi}^{r}, \phi_{s}\}_{PB} = i\delta_{r}^{s}$ and construct new operators which satisfy an algebra inherited from the gauge group
\begin{equation}
	L^{S} = \tilde{\phi}^{r}(T^{S})_{r}{}^{s}\phi_{s}; \qquad \left\{L^{S}, L^{T}\right\}_{PB} = f^{STU}L^{U}
	\label{als}
	\end{equation}
and will consequently work just as well to represent the coupling between the matter field and the gauge field. We incorporate these fields into the worldline action in the same way as in \cite{JOdof}, which leads to an action similar to (\ref{worldlinephi}):
\begin{equation}
	 S[\omega, \psi, \tilde{\phi}, \phi] = \int_{0}^{2\pi} d\tau \left[ \frac{\dot{\omega}^{2}}{2T} + \frac{i}{2}\psi \cdot \dot{\psi} + i\tilde{\phi}^{r}  \dot{\phi}_{r} + \tilde{\phi}^{r} \mathcal{A}_{r}{}^{s} \phi_{s}\right],
	 	 \label{worldlinephiB}
\end{equation}
that is now based upon bosonic colour fields. As discussed previously in section \ref{secStruct}, it is also possible to write down a globally supersymmetric version of this worldline theory by introducing super-partners (which would now need to be fermionic\footnote{This change in the nature of the components of the colour multiplet also means that the variations of $\tilde{z}$ and $z$ in (\ref{susycz}) lose the minus sign.}) to $\tilde{\phi}$ and $\phi$, which could then be gauged if desired. For brevity we omit further discussion on this point as it will not be important for what is to come. However, it is instructive to consider the Fock space of the commuting variables so we briefly recap the main points discussed in \cite{JOdof}.

We promote the $\tilde{\phi}^{r}$ and $\phi_{s}$ to creation and annihilation operators which in the present case span an infinite dimensional space $\{\state{n_{r}}\}$ for $ n_{r}\in \mathbb{N}$ for each index $r$. The fundamental commutation relations inherited from the Poisson brackets are $[\hat{\phi}^{\dagger r}, \hat{\phi}_{s}] = -\delta_{s}^{r}$ which are solved by taking $\hat{\phi}^{\dagger r} = \tilde{\phi}^{r}$ and $\hat{\phi}_{r}= \partial_{\tilde{\phi}^{r}}$ when acting on wave-functions $\Phi(x, \tilde{\phi})$ which have an expansion
\begin{equation}
	\Phi(x, \tilde{\phi}) = \Phi(x) + \tilde{\phi}^{r_{1}} \Phi_{r_{1}}(x) + \tilde{\phi}^{r_{1}}\tilde{\phi}^{r_{2}} \Phi_{r_{1}r_{2}}(x) + \ldots
	\label{Focks}
\end{equation}
Notice that the commuting nature of the variables means that the Fock space is infinite dimensional so that the expansion above does not terminate. The sum is over all components whose indices transform as \textit{symmetric} tensor products of the fundamental representation so $\Phi(x, \tilde{\phi})$ is described by a reducible representation of the gauge group. We solve this problem as in the previous case: the number operator for these fields is $\hat{n} = \hat{\phi}^{\dagger r}\phi_{r}$ which one can use to project onto states with the correct occupation number. To do so we introduce the constraint $\left(\hat{n} - n\right)\phy = 0$. This acts on wave-functions as $\left(\tilde{\phi} \cdot \partial_{\tilde{\phi}} - n\right)\Phi(x, \tilde{\phi}) = 0$ to pick out the component with $n$ symmetrised indices. From the point of view of the path integral this constraint can again be enforced by gauging the $U(1)$ symmetry of the colour fields \cite{JOdof}, achieved by modifying the functional integration over $\tilde{\phi}$ and $\phi$ to 
\begin{equation}
	\int \mathscr{D}\tilde{\phi}\mathscr{D}\phi \mathscr{D} a\, e^{ i\int d\tau \, (n +\frac{N}{2}) a(\tau) -  a(\tau) \tilde{\phi}^{r} \phi_{r}}.
	\label{intaB}
\end{equation}
Under this symmetry the fields transform as $\delta_{\vartheta}\tilde{\phi} = i\vartheta\tilde{\phi}$, $\delta_{\vartheta}\phi = -i\vartheta \phi$ and $\delta_{\vartheta}a = \dot{\vartheta}$, where $\vartheta$ is an infinitesimal real parameter. As before, we have introduced the topological term $\int d\tau \left(n + \frac{N}{2} \right)a(\tau)$ as a Chern-Simons coupling for this gauge field in order to impose the desired constraint (the operator ordering ambiguity that arises upon quantisation is this time resolved by symmetrising which leads to the plus sign). We wish to show now that this additional $U(1)$ symmetry is sufficient to project away the unwanted states in the decomposition of $\Phi$ in (\ref{Focks}) so as to pick out a single wavefunction component that transforms irreducibly. We shall do this by functionally quantising the colour fields and the $U(1)$ field. 

\subsection{Functional Quantisation}
\label{secFuncB}
The partition function associated to these new fields is (we continue to carry out path integrals in Euclidean space) given by
\begin{equation}
	\mathcal{Z} = \int \frac{\mathscr{D}a\mathscr{D}\tilde{\phi}\mathscr{D}\phi}{\textrm{Vol(U(1))}} e^{-  \int_{0}^{2\pi} d\tau \left[ \tilde{\phi}^{r} (\delta_{r}^{s}D - i\mathcal{A}_{r}{}^{s})\phi_{s} - i(n + \frac{N}{2})a\right]},
\end{equation}
where $D = \left(\frac{d}{d \tau} + i a\right)$, which represents the complete coupling between the matter field and the gauge field, implemented by the auxiliary colour fields. There are important differences, however, due to the bosonic nature of the colour fields, which will lead to an interesting result which is of importance for remaining sections. We proceed as before by fixing the the $U(1)$ symmetry which we do by setting $2\pi a(\tau) = \theta$, where $\theta$ is a constant angle in $[0, 2\pi]$. This angle is the $U(1)$ modulus which parameterises inequivalent Wilson-loops of the worldline gauge field. The Faddeev-Popov determinant associated with this fixing remains trivial so with this gauge choice we arrive at
\begin{equation}
	\mathcal{Z} = \int_{0}^{2\pi}\frac{d \theta}{2\pi} e^{i (n+ \frac{N}{2})\theta}\mathcal{Z}\left[\mathcal{A}, \theta\right],
\end{equation}
where the partition function of the colour fields continues to be denoted by
\begin{equation}
	\mathcal{Z}\left[\mathcal{A}, \theta\right] = \int_{\textrm{PBC}}\!\!\!\!\mathscr{D}\tilde{\phi} \mathscr{D}\phi\, e^{- \int_{0}^{2\pi} d\tau\, \tilde{\phi}^{r}(\delta_{r}^{s}D + \mathcal{A}_{r}{}^{s}) \phi_{s}},
	\label{Zas}
\end{equation}
with periodic boundary conditions ($\textrm{PBC}$) on the colour fields. As above it is then possible to absorb the coupling to the gauge field into these fields by using the field re-definition
\begin{align}
	\tilde{\phi}\left(\tau\right) &\rightarrow \tilde{\phi}\left(\tau\right)\exp{\left(-i \int_{0}^{\tau} a\left(\tau^{\prime}\right)d\tau^{\prime}\right)} = \tilde{\phi}(\tau)e^{- \frac{i\theta}{2\pi} \tau}\nonumber \\
	 \phi\left(\tau \right) &\rightarrow \exp{\left(i \int_{0}^{\tau} a\left(\tau^{\prime}\right)d\tau^{\prime}\right)}\phi\left(\tau\right) = e^{\frac{i\theta}{2\pi} \tau}\phi(\tau),
	 \label{redefB}
\end{align}
which implies the twisted boundary conditions (TBC) $\tilde{\phi}(2\pi) = e^{i\theta}\tilde{\phi}(0)$ and $\phi(2\pi) = e^{-i\theta}\phi(0)$ but removes the awkward $\theta$-dependence from the action. Integrating over these fields then gives a functional determinant
	\begin{align}
	Z\left[\mathcal{A}\right] &= \frac{1}{\underset{\textrm{\tiny TBC}}{\det}{\left(i\left(\frac{d}{d\tau} -i \mathcal{A}\right)\right)}},
	\label{Zdets}
	\end{align}
which we find as above through the product of the eigenvalues of the operator in (\ref{Zdets}). These eigenvalues can again be expressed in terms of eigenvalues of the Wilson-loop, which we denote by $\rho$, and a simple calculation similar shows that the determinant is proportional to\footnote{The difference in the current setting is that the eigenfunctions must obey periodic boundary conditions. Denoting the eigenvalues of the operator in (\ref{Zdets}) by $\mu_{+}$, these eigenfunctions can be written as $v(\tau) = g(\tau)v(0)e^{-i\mu_{+} \tau}$ and we require the $\mu_{+}$ be related to the eigenvalues of $g(2\pi)$ by
\begin{equation}
	g\left(2\pi\right)v\left(0\right) = \rho v\left(0\right); \qquad \rho e^{-2\pi i \mu_{+}} = e^{i \theta}.
\end{equation}
The solution to this condition is $\mu_{+} = n  + \frac{\ln{\left(\rho e^{i\theta}\right)}}{2\pi i}$.}
	\begin{align}
	\prod_{\{\rho\}}\prod_{n > 0}\left(1 - \frac{\left(\ln{(\rho e^{i\theta})}\right)^{2}}{\left(  2n \pi i\right)^{2}}\right) \ln{(\rho e^{i\theta})},
	\label{prodss}
\end{align}
which reminds one of the infinite product expansion of the $\sin$ function (we have paired positive and negative integers together which is part of our regularisation procedure). Indeed, taking the product over $\rho$ and inverting the result we arrive at 
	\begin{align}
	Z\left[\mathcal{A}, \theta\right] \propto \left(\det{ \left(\sqrt{e^{i\theta}g\left(2\pi\right)} - 1/\sqrt{e^{i\theta}g\left(2\pi\right)}\right)}\right)^{-1}.
	\label{detss}
\end{align}
which can be compared with (\ref{dets}). This result justifies the introduction of the bosonic auxiliary fields and gives a concrete realisation of their purpose in generating objects related to the Wilson-loop providing the interaction between matter and the gauge field. The determinant in rounded brackets depends in a non-trivial way on the Wilson-loop. We have shown in section \ref{secFunc} (see also \cite{Me1}) that it can be re-expressed in terms of group invariant objects built from traces of $g(2\pi)$ in fully anti-symmetric representations and we will use this extensively in what follows.

To proceed it is necessary to expand (\ref{detss}) as an infinite series. We should indeed expect to encounter such an infinite sum since the Fock space of the $\tilde{\phi}$, $\phi$ theory is infinite dimensional. From the series expansion of (\ref{detss}) we must arrange the resulting products of traces in anti-symmetric representations into traces of the Wilson-loop taken in \textit{symmetric} representations as anticipated from (\ref{Focks}). We will then make use of the $U(1)$ field to project onto a chosen symmetric representation by carrying out the integral over this modulus. As in section \ref{secFunc}, we will make use of the notation based upon Young Tableaux, since this allows us to make statements that hold for arbitrary $N$.

The change of sign in (\ref{detss}) means that the result given in (\ref{SUN}) must also be modified by a few changes of sign: we find \Yvcentermath1
\begin{align}
	\mathcal{Z}_{\mathbf{N}}	\left[\mathcal{A}, \theta\right] &\propto \bigg( \tr g({ \, \boldsymbol{\cdot} \,})e^{\frac{N}{2} i \theta} - \mathrm{tr} g({\,\Tiny\yng(1)\, } )e^{\frac{N-2}{2} i \theta} + \mathrm{tr}g({\Yvcentermath1\,\Tiny\yng(1,1)\, } )e^ {\frac{N-4}{2}i \theta} + \ldots \nonumber \\
	&\qquad \qquad\qquad +(-1)^{N-1}\tr g( \underset{ \,\Tiny\yng(1,1)\, }{\overset{ \,\Tiny\yng(1,1)\, } {\rvdots} } )e^{-\frac{N-2}{2} i \theta} + (-1)^{N} \tr g({ \,\boldsymbol{\cdot}  \,})e^{-\frac{N}{2} i \theta}\bigg)^{-1}.
	\label{SUNs}	
	\Yvcentermath0
\end{align} 
To project onto a certain representation we must integrate this over the $U(1)$ modulus with associated Chern-Simons term $\exp{\left( i(n + \frac{N}{2}) \right)}$. It is convenient to cancel the leading $e^{-i\frac{N}{2}}$ by factorising $e^{i\frac{N}{2}}$ out of the right hand side of (\ref{SUNs}). Doing so leaves an integral representation for $\mathcal{Z}$ \Yvcentermath1
\begin{align}
	\int_{0}^{2\pi} \frac{d\theta}{2\pi}\, e^{i n\theta } \bigg(1 - \mathrm{tr} g( {\,\Tiny\yng(1)\, } )e^{-i \theta} &+ \mathrm{tr} g({\Yvcentermath1 \,\Tiny\yng(1,1)\, } )e^ {-2 i \theta} + \ldots + (-1)^{p+1}\tr g(\underset{ \,\Tiny\yng(1,1)\, }{\overset{ \,\Tiny\yng(1,1)\, } {\rvdots} })e^{-p i \theta}+\ldots \nonumber \\
	&\quad\qquad + (-1)^{-(N-1)}\tr g(\underset{ \,\Tiny\yng(1,1)\, }{\overset{ \,\Tiny\yng(1,1)\, } {\rvdots} } )e^{(N -1) i \theta} + (-1)^{N}\tr g( { \,\boldsymbol{\cdot}  \,})e^{-i N\theta}\bigg)^{-1}.
	\label{projects}
	\Yvcentermath0
\end{align}
A binomial expansion of the rounded bracket produces a non-terminating series which can be arranged in increasing powers of $e^{-i\theta}$; the integration over this variable then picks out the coefficient of the term of order $e^{-i n \theta}$. We illustrate this by considering the first few terms in the expansion of (\ref{projects}):\Yvcentermath1
\begin{align}\Yvcentermath1
	\int_{0}^{2\pi} \frac{d\theta}{2\pi}\,  e^{i n\theta } \bigg[1 \!&+\!  \mathrm{tr} g( {\Yvcentermath1\,\Tiny\yng(1)\, }) e^{-i \theta}\! +\! \bigg((\tr  g( {\Yvcentermath1\,\Tiny\yng(1)\, }))^{2}-\mathrm{tr} g({\Yvcentermath1\,\Tiny\yng(1,1)\, })\bigg) e^ {-2 i \theta}\! +\! \bigg((\tr g( {\,\Tiny\yng(1)\, }))^{3}-2\mathrm{tr} g({\,\Tiny\yng(1,1)\, })\mathrm{tr} g({\,\Tiny\yng(1)\, }) \nonumber\\
	&  + \mathrm{tr} g({\,\Tiny\yng(1,1,1)\, })\bigg) e^ {-3 i \theta} \! + \!\bigg((\mathrm{tr} g({\,\Tiny\yng(1)\, }))^4 + (\mathrm{tr} g({\,\Tiny\yng(1,1)\, }))^2 + 2\mathrm{tr} g({\,\Tiny\yng(1,1,1)\, }) \mathrm{tr} g({\,\Tiny\yng(1)\, }) \nonumber \\
	&- 3 (\mathrm{tr} g({\,\Tiny\yng(1)\, }))^2 \mathrm{tr} g({\,\Tiny\yng(1,1)\, })
	 - \mathrm{tr} g({\,\Tiny\yng(1,1,1,1)\, })\bigg)e^{-4i\theta} + \mathcal{O}\left(e^{-5i\theta}\right)\bigg]\,.
	\label{series}
	\Yvcentermath0
\end{align}
Setting $n = 0$ or $n = 1$ clearly provides the expected trace of the Wilson-loop in the representations with $n$ indices. After this some group theory identities are uncovered. According to our analysis of the Hilbert space, setting $n = 2$ should provide the trace of $g(2\pi)$ in the representation with two fully symmetrised indices. The Young Tableau of this representation is ${\Yvcentermath1 \Tiny \yng(2)\Yvcentermath0}$ which has dimension $\frac{N}{2}(N+1)$. So (\ref{series}) implies that
\begin{equation}
\Yvcentermath1
	\tr g({\,\Tiny\yng(2)\, }) \propto (\tr g({\,\Tiny\yng(1)\, }))^2 - \tr g({\,\Tiny\yng(1,1)\, }),
\Yvcentermath0
\end{equation}
and it is easy to check that the dimensions of the representations on the right hand side combine correctly to agree with the dimension of the left hand side. Similarly, by choosing $n = 3$ we should find the trace of the Wilson-loop transforming in the $\frac{N}{6}(N+1)(N+2)$ dimensional representation ${\Yvcentermath1\,\Tiny\yng(3)\,\Yvcentermath0 }$, which transforms as the symmetric product of three copies of the fundamental. Integrating (\ref{series}) over $\theta$ then yields
\begin{equation}
\Yvcentermath1
	\tr g( {\,\Tiny\yng(3)\, } )\propto (\tr g({\,\Tiny\yng(1)\, }))^{3}-2\mathrm{tr}g({\,\Tiny\yng(1,1)\, })\mathrm{tr}g({\,\Tiny\yng(1)\, }) + \mathrm{tr}g({\,\Tiny\yng(1,1,1)\, }).
\Yvcentermath0
\end{equation}
A straightforward calculation shows that the dimensions of the representations of the right hand side are consistent, since $N^{3} - 2\times \frac{N}{2}(N-1) + \frac{N}{6}(N-1)(N-2)$ is indeed the dimension of the representation on the left hand side. Finally, by putting $n = 4$ and carrying out the integration another identity is found:
\begin{equation}
\Yvcentermath1
	\tr g({\,\Tiny\yng(4)\, }) \propto (\mathrm{tr}g({\,\Tiny\yng(1)\, }))^4 + (\mathrm{tr}g({\,\Tiny\yng(1,1)\, }))^2 + 2\mathrm{tr}g({\,\Tiny\yng(1,1,1)\, }) \mathrm{tr}g({\,\Tiny\yng(1)\, }) - 3 (\mathrm{tr}g({\,\Tiny\yng(1)\, }))^2 \mathrm{tr}g({\,\Tiny\yng(1,1)\, }) - \mathrm{tr}g({\,\Tiny\yng(1,1,1,1)\, })
\Yvcentermath0
\end{equation}
which relates the traces of $g(2\pi)$ in anti-symmetric representations to that in the four index totally symmetric representation. 

These identities are of course related to the decomposition of tensor products of the fundamental representation into the direct sum of irreducible representations. For example, the decompositions
\begin{equation}
\Yvcentermath1
	{\tiny\yng(1) } \otimes {\tiny\yng(1) } = {\tiny\yng(2) } \oplus {\tiny\yng(1,1)} \qquad \textrm{and} \qquad {\tiny\yng(1) } \otimes {\tiny\yng(1)} \otimes {\tiny\yng(1)} = {\tiny\yng(1,1,1) } \oplus 2{\,\tiny\yng(2,1) }  \oplus {\Yvcentermath1 \tiny\yng(3)}
\Yvcentermath0
\label{young}
\end{equation}
take a similar form to the first two identities presented above (the difference being that the identities are related to the \textit{traces of the Wilson-loop} in the representations signified by the Young Tableaux). Furthermore a fourth tensor product with the fundamental gives
\begin{equation}
\Yvcentermath1
	{\tiny\yng(1) } \otimes {\tiny\yng(1)} \otimes {\tiny\yng(1)} \otimes {\tiny\yng(1)} = {\tiny\yng(1,1,1,1)}\oplus 3{\,\tiny\yng(2,1,1)}\oplus {\tiny\yng(2,2)} \oplus 3{\,\tiny\yng(3,1)} \oplus {\tiny\yng(4)},
\Yvcentermath0
\end{equation}
into which form (in combination with the obvious tensor product decompositions) the third identity quoted above can be cast. The three identities we have presented here, and their generalisations derived from higher order terms in (\ref{series}), are easily verified in the simple case that the gauge group is $SU(3)$ or $SU(5)$ as we have previously considered, if one uses a global $SU(N)$ transformation to rotate $g(2\pi)$ onto the Cartan subalgebra, determines its eigenvalues in the appropriate representations and combines them for the traces given in the equations above. Straightforward algebra proves, for example, that for $SU(3)$ the traces of the Wilson-loop are related as expected:
\begin{align}
	\tr(g_\mathbf{6}) &= (\tr(g_\mathbf{3}))^2 - \tr(g_\mathbf{\bar{3}})\nonumber \\
	\tr(g_\mathbf{10}) &= (\tr(g_\mathbf{3}))^3 - 2\tr(g_\mathbf{3})\tr(g_\mathbf{\bar{3}}) + 1 \nonumber \\
	\tr(g_\mathbf{15'}) &= (\tr(g_\mathbf{3}))^4 + (\tr(g_\mathbf{\bar{3}}))^2 + 2\tr(g_\mathbf{3}) - 3(\tr(g_\mathbf{3}))^2\tr(g_\mathbf{\bar{3}}).
\end{align}  
Similar results hold in the case of $SU(5)$ and other Lie groups. 

To lend further weight to the claim that the terms in (\ref{series}) arrange themselves into traces of the super-Wilson loop in the fully symmetric representations it is informative to count the dimensions of the representations that combine to make each term. The trace of the Wilson-loop is equal to the sum of its eigenvalues, whose number match the dimension of the representation in which it transforms. We use this to count the number of terms making up each trace in (\ref{projects}). Since the dimension of the representation of $SU(N)$ with $k$ fully antisymmetrised indices is $\begin{pmatrix}N\\k\end{pmatrix}$ the number of terms building each trace in the denominator of (\ref{projects}) is
\begin{align}
	\bigg[\nCr{N}{0} - \nCr{N}{1}e^{-i \theta} &+ \nCr{N}{2} e^ {-2 i \theta} + \ldots + (-1)^{p}\nCr{N}{p}e^{-p i \theta}+\ldots \nonumber \\
	&\quad+ (-1)^{N-1}\nCr{N}{N\!-\!1}e^{-(N -1) i \theta} + (-1)^{N}\nCr{N}{N}e^{-i N\theta}\bigg] = \left(1 - e^{-i \theta}\right)^{N}
	\label{dims}
\end{align}
where we have made use of the binomial theorem to rewrite the series as the difference of two terms. This trick will prove enough to extract the dimension of the representation which makes up the coefficient of $e^{-p i \theta}$ in (\ref{series}). To understand how, it suffices to invert the above equation and use the binomial expansion once again to arrive formally at
\begin{equation}
	\left(1 - e^{-i \theta}\right)^{-N} = \sum_{p = 0}^{\infty} \nCr{\!-\!N}{p} (-1)^{p}e^{-p i \theta}.
	\label{dimsinvert}
\end{equation}
The binomial coefficients for negative powers are defined by
\begin{equation}
	\nCr{\!-\!N}{p} = \frac{-N(-N-1)\ldots (-N-(p-1))}{p!} = (-1)^{p}\nCr{N\!+\!p\!-\!1}{p},
\end{equation}
which when substituted into (\ref{dimsinvert}) produces
\begin{equation}
	\left(1 - e^{-i \theta}\right)^{-N} =\sum_{p = 0}^{\infty} \nCr{N\!+\!p\!-\!1}{p} e^{-p i \theta}.
\end{equation}
This equation shows that the coefficient of the term in (\ref{series}) involving $e^{-p i \theta}$ is made up of a sum containing exactly $\nCr{N\!+\!p\!-\!1}{p}$ terms involving the eigenvalues of the Wilson loop in various representations. However, this is the dimension of the representation of $SU(N)$ which has $p$ fully symmetric indices, and the trace of $g(2\pi)$ in that representation is the sum over this number of eigenvalues. There may of course be other representations which have the same dimension, but the analysis of the Fock space of colour carrying fields in section \ref{secSymm} shows that it will be the fully symmetric representations that are involved. This allows us to conclude that (\ref{projects}) and (\ref{series}) can be written more informatively as\Yvcentermath1
\begin{align}
	\mathcal{Z} = \int_{0}^{2\pi} \frac{d\theta}{2\pi}\,  e^{i n\theta } \bigg(\tr g({ \,\boldsymbol{\cdot}  \,}) &+ \mathrm{tr} g( {\,\Tiny\yng(1)\, } )e^{-i \theta}+ \mathrm{tr}g({\Yvcentermath1 \,\Tiny\yng(2)\, }) e^ {-2 i \theta} +\mathrm{tr}g({\Yvcentermath1 \,\Tiny\yng(3)\, }) e^ {-3 i \theta} +\ldots  \nonumber \\
	&+\tr g( { \,\Tiny\yng(2)}\!\!  \rhdots \!\! {\Tiny\yng(2)\, } ) e^{-p i \theta} + \ldots \bigg),
	\label{seriess}\Yvcentermath0
\end{align}
containing an infinite sum over traces of the super-Wilson loop in all symmetric representations, each carrying a $U(1)$ hypercharge. The integral over $\theta$ then selects from this sum the single representation which has $n$ indices. 

This concludes the discussion of one method for generating the Wilson-loop coupling between a gauge field and a spin 1/2 matter field which transforms in an arbitrary totally symmetric representation of the symmetry group. This is an interesting problem for the worldline formalism of quantum field theory since it allows for a simple description of such matter fields in a framework which enjoys significant calculational advantages and has met great success in perturbative calculations (see for example \cite{Bastwl2}). However, up until now we have been restricted to projecting onto fully symmetric representations, which is a limitation that can be overcome. We now explain how to use this as the basis with which an arbitrary irreducible representation can be built up by introducing families of the commuting colour fields and a method of projecting out unwanted contributions to the result. 
\section{Mixed symmetry tensors with bosonic colour fields}
\label{secMixed2}
To generalise these ideas to a theory describing matter fields which transform in an arbitrary representation of the gauge group it is useful to follow the same ideas as in section \ref{secMixed} where the colour fields were anti-commuting (as we originally advocated in \cite{JOdof}). As we have shown, this allowed us to build wavefunctions out of components transforming in tensor products of anti-symmetric representations. In the present case, however, a specified representation will be constructed by taking tensor products of the fully symmetric representations uncovered in the previous section. To do this we repeat the process of introducing multiple families of bosonic fields. Each family will produce the trace of the super-Wilson loop in an individually chosen symmetric representation and the families will be combined to generate products of these traces. These must be manipulated into a single expression which transforms in the desired irreducible representation. 

As before the generalisation of the partition function is arrived at by extending it to include $F$ copies of the bosonic variables. We continue to denote these colour fields by $\tilde{\phi}^{r}_{k}$ and $\phi_{kr}$, where $k$ indexes the family of the field. In Minkowski space the dynamics are given by
\begin{equation}
	 S[\omega, \psi, \tilde{\phi}, \phi] = \int_{0}^{2\pi} d\tau \left[ \frac{\dot{\omega}^{2}}{2T} + \frac{i}{2}\psi \cdot \dot{\psi} + \tilde{\phi}^{r}_{k}(i\delta_{s}^{r}D_{kj} + \mathcal{A}_{r}{}^{s}\delta_{kj}) \phi_{js} \right],
	 \label{Sf}
\end{equation} 
where $D_{kj} = (\delta_{kj}\frac{d}{d\tau} + ia_{kj})$ is the covariant derivative constructed out of gauge fields $a_{kj}(\tau)$. These fields supply the worldline theory with a $U(F)$ symmetry which generalises the $U(1)$ symmetry that appeared in the previous section. Under a transformation, generated by the infinitesimal parameters $\alpha^{k}_{l}$ the fields vary as
\begin{equation}
	\delta_{\alpha}\tilde{\phi}^{r}_{k} = i \tilde{\phi}^{r}_{l}\alpha_{lk}; \qquad \delta_{\alpha}\phi_{kr} =  -i \alpha_{kl}\phi_{lr}; \qquad \delta_{\alpha}a_{kj} = \dot{\alpha}_{kj} - i[\alpha, a]_{kj},
	\label{Uf}
\end{equation}
and are generated by the conserved currents which are now constructed out of bosonic constituents as $N_{kj} = \tilde{\phi}_{k}^{r}\phi_{jr}$. We have shown in \cite{JOdof} that, as it stands, the partition function for the colour fields supplies a Hilbert space in which the wavefunction components transform in tensor products of fully symmetric representations. As in \cite{JOdof}, and the previous section, it is necessary to partially gauge the $U(F)$ symmetry in order to project onto a single irreducible representation. To retain the ability to add independent Chern-Simons terms for each family one must introduce $a_{kj}$ only for $k\geqslant j$, corresponding to gauging the first class subgroup generated by the corresponding $N_{kj}$. This means that the diagonal gauge fields transform in an Abelian way so that we can construct Chern-Simons terms 
\begin{equation}
	S[a] = \int d\tau \, \sum_{k=1}^{F} s_{k} a_{kk}(\tau)
\end{equation}
which are invariant under the transformations in (\ref{Uf}). The equations of motion of the $a_{kk}$ will then impose the constraints which pick out the wavefunction component which transforms as the $F$-fold tensor product with $n_{k}$ fully symmetric indices in the $k^{th}$ family \cite{JOdof}. We denote the set of integers specifying the occupation numbers of each family by $n = (n_{1}, n_{2}, \ldots , n_{F})$, taking $n_{1} \leqslant n_{2} \leqslant \ldots \leqslant n_{F}$ which enter into the Chern-Simons levels $s_{k}$. The off-diagonal gauge fields impose further conditions on the symmetries that exist over indices belonging to different families. The only wavefunction component that survives lies in the kernel of all of the $N_{kj}$ for $k > j$. The action of these constraints on the wavefunction components mirror the way that elements of the Lie algebra are combined in the tensor product and ensure that unwanted representations cannot enter the path integral at intermediate steps.

\subsection{Gauge fixing and evaluation of the path integral}
We are once again interested in calculating the path integral over gauge inequivalent configurations of the dynamical fields and the colour fields. Focusing for the moment on the colour fields, we compute
\begin{align}
	\mathcal{Z}^{(F)} = \int \frac{\mathscr{D}\tilde{\phi}\mathscr{D}\phi\mathscr{D}a}{\textrm{Vol(Gauge)}} \exp\bigg(-\int_{0}^{2\pi} d\tau \bigg[ \tilde{\phi}^{r}_{k} \big(\delta_{r}^{s}\frac{d}{d\tau} &- i\mathcal{A}_{r}{}^{s}\big)\phi_{ks} \nonumber \\
	& +i\textstyle{\sum\limits_{k=1}^{F}}a_{k}\left(N_{k} - s_k\right)+i \textstyle{\sum\limits_{j < k}}a_{kj}N_{kj}\bigg]\bigg)
\end{align}
As is by now familiar we must first gauge fix the auxiliary $U(F)$ fields in order to deal with the quotient by volume of the auxiliary symmetry group and we choose the $a_{kj}$ to be constants:
\begin{equation}
	2\pi \hat{a}_{kj} = \begin{pmatrix}
		\theta_{1} & 0 & \cdot & 0 \\ 0 & \theta_{2} & \cdot & 0 \\ \cdot & \cdot & \cdot & \cdot \\ 0 & 0 & \cdot & \theta_{F}
	\end{pmatrix}
	\label{afixB}
\end{equation}
where the diagonal elements are the angular moduli that we must integrate over. The Faddeev-Popov determinant associated to this gauge fixing gives the same modular measure as in section \ref{secMixed}
\begin{equation}
	\mu\left(\{\theta_{k}\}\right) = \prod_{j<k}\mu\left(\{\theta_{k}, \theta_{j}\}\right)= \prod_{j< k}2i \sin{\left(\frac{\theta_{j} - \theta_{k}}{2}\right)}.
\end{equation}
Now, on our chosen gauge slice the partition function of the colour fields becomes
\begin{equation}
	\mathcal{Z}^{(F)}\left[\mathcal{A}, \{\theta_{k} \}\right] = \prod_{k=1}^{F}\int_{\textrm{PBC}}\!\!\!\!\mathscr{D}\tilde{\phi_{k}} \mathscr{D}\phi_{k}\, e^{-\int_{0}^{2\pi} d\tau\, \tilde{\phi}^{r}_{k}(\delta_{r}^{s}D_{k} -i \mathcal{A}_{rs}) \phi_{ks}},
	\label{ZasF}
\end{equation}
where each family of bosonic fields has period boundary conditions and the covariant derivative is diagonal so $D_{k} = (\frac{d}{d\tau} + \frac{i\theta_{k}}{2\pi})$. When we then return to integrate this partition function against the Chern-Simons moduli we will include the factors which fix the occupation numbers in each family and the measure which ensures irreducibility.  It is easy to see that the path integral factorises into a product of inverse functional determinants, where the factor corresponding to each family carries an independent hypercharge:
\begin{align}
	Z^{(F)}\left[\mathcal{A}, \{\theta_{k}\}\right] \propto \prod_{k=1}^{F} \left(\det{ \left(\sqrt{e^{i\theta_{k}}g\left(2\pi\right)} - 1/\sqrt{e^{i\theta_{k}}g\left(2\pi\right)}\right)}\right)^{-1}.
\label{detssF}
\end{align}
As we have shown, each factor in the product turns into an infinite series involving the traces of the super-Wilson loop in all fully symmetric representations with an angular exponent related to the number of indices which specify how the representation transforms. Reinstating the functional integrals over the matter degrees of freedom and the remaining integrals over the $U(F)$ moduli, the complete worldline theory is now given by
\begin{align}
	\int_{0}^{\infty} \frac{dT}{T} \oint \mathscr{D}\omega \mathscr{D}\psi\, e^{-\frac{1}{2} \int_{0}^{2\pi} \frac{\dot{\omega}^{2}} {T} + \psi \cdot \dot{\psi} \,d\tau }\prod_{k = 1}^{F}\int_{0}^{2\pi} \frac{d\theta_{k}}{2\pi}\, e^{is_{k}\theta_{k} } \mu\left(\{\theta_{k}\}\right) \mathcal{Z}^{(F)}\left[\mathcal{A},\{\theta_{k}\}\right]
	\label{ZsF}
\end{align}
which the crucial factor being the measure which is required to impose irreducibility. 

With these computations, the worldline theory (\ref{ZsF}) leads to a general formula which generates the trace of the Wilson loop in a single irreducible representation, extending the projection presented in (\ref{seriess}) to allow the matter field to transform in an arbitrary representation. By following the same straightforward algebra mentioned in section \ref{secMixed} we can write the resulting formula compactly as
\begin{align}\Yvcentermath1
	&\int_{0}^{\infty} \!\frac{dT}{T} \!\oint \!\mathscr{D}\omega \mathscr{D}\psi\, e^{-\frac{1}{2} \int_{0}^{2\pi} \frac{\dot{\omega}^{2}} {T} + \psi \cdot \dot{\psi} \,d\tau}\prod_{k = 1}^{F}\int_{0}^{2\pi} \!\frac{d\theta_{k}}{2\pi}\, e^{i n_{k} \theta_{k} } \prod_{j<k}\left(1\! -\! e^{- i \theta_{j}}e^{i \theta_{k}}\right) \times \nonumber \\
	& \prod_{k = 1}^{F} \bigg(\tr g({ \,\boldsymbol{\cdot}  \,}) +\mathrm{tr} g( {\,\Tiny\yng(1)\, } )e^{-i \theta_{k}}+ \mathrm{tr}g({\Yvcentermath1 \,\Tiny\yng(2)\, }) e^ {-2 i \theta_{k}} +\ldots  + \tr g( { \,\Tiny\yng(2)}\!\!  \rhdots \!\! {\Tiny\yng(2)\, } ) e^{-p i \theta_{k}} + \ldots \bigg). 
	\Yvcentermath0
	\label{ZF''}
\end{align}
In this equation we have chosen to use the results of section \ref{secSymm} to write the infinite sum in (\ref{series}) in terms of the traces of the super-Wilson loop in the fully symmetric representations; the general term involves the trace taken in the representation with $p$ fully symmetric indices. We have also made the occupation numbers explicit having absorbed part of the Chern-Simons factors into the product at the end of the top line. The mechanics of this formula are as follows: the $k$th family produces the trace of $g(2\pi)$ in the representation with $n_{k}$ fully symmetric indices. The measure is responsible for combining these traces in such a way as to form the trace in the tensor product of these representations and to subtract the traces in all the unwanted representations in the tensor product decomposition. As we have seen, the integrals over the modular parameters pick out a single contribution from each of the terms in the product on the bottom line.

As we explained in the case of fermionic colour fields, the idea is that various contributions are combined together so as to eventually leave only the trace of the super-Wilson loop in the representation whose Young Tableau this time has $n_{k}$ columns in each row. So, for example, taking $F = 6$ and $n = (1,2,4,4,5,8)$, the integration over the $U(6)$ moduli produces a worldline path integral
\begin{equation}
	\int_{0}^{\infty} \frac{dT}{T} \oint \mathscr{D}\omega \mathscr{D}\psi\, e^{-\frac{1}{2} \int_{0}^{2\pi} \frac{\dot{\omega}^{2}} {T} + \psi \cdot \dot{\psi} \, d\tau } \,\tr_{R} \mathscr{P} \exp{\left(i\int_{0}^{2\pi} \mathcal{A}[\omega(\tau), \psi(\tau)] d\tau\right)}
\end{equation}
where the representation, $R$, in which the super-Wilson loop is taken to transform has a Young Tableau with six rows and $n_{k}$ columns in each row:
\begin{equation}\Yvcentermath1
	{\,\tiny \yng(8,5,4,4,2,1)\, }
	\Yvcentermath0
\end{equation}
as desired. In this instance it remains possible to project onto the adjoint representation of the symmetry group. Now we need $F = N - 1$ families and set $n = (1, 1, \ldots, 2)$, which projects onto the irreducible representation with Young Tableau
\begin{equation}\Yvcentermath1
	 \underset{ \,\tiny\yng(1,1)\, }{\overset{\hphantom{\tiny\yng(1)} \,\tiny\yng(2,1)\, } {\rvdots}} \, .
	\Yvcentermath0
\end{equation}
The first column of this diagram has $N-1$ boxes so the dimension of the representation is $N^{2} - 1$ as required. It proves convenient to introduce the worldline Wilson-loop variables $z_{k} = e^{i \theta_{k}}$, as we did in the fermionic case, so that we may write a compact version of (\ref{ZF''}) in the same notation as \cite{JOdof}:
\begin{align}
		\int_{0}^{\infty} \!\frac{dT}{T} \!\oint \!\mathscr{D}\omega \mathscr{D}\psi\, e^{-\frac{1}{2}\int \frac{\dot{\omega}^{2}} {T} + \psi \cdot \dot{\psi} \,d\tau}\prod_{k = 1}^{F}\oint \!\frac{dz_{k}}{2\pi i} \prod_{l<j}\left(1 - \frac{z_{j}}{z_{l}}\right)  \prod_{k= 1}^{F} \sum_{p_{k} = 1}^{N} \frac{\tr g(2\pi)_{(p_{k})}}{z_{k}^{p_{k}+1-n_{k}}}
	\Yvcentermath0
	\label{ZF''z}
\end{align}
where the integrals with respect to the complex parameters $z_{k}$ are closed loops winding around the origin. In (\ref{ZF''z}), the notation $\tr g(2\pi)_{(p_{k})}$ now means the trace of the super-Wilson loop in the representation with $p_{k}$ fully symmetric indices. 

To demonstrate how the modular measure imposes irreducibility, and to compare its behaviour to the case we have previously discussed, we provide an example calculation involving $F = 3$ families, taking (as we did earlier) $n = (1,1,2)$. 
\begin{align}\Yvcentermath1
	\tr g( {\,\Tiny\yng(2,1,1)\, } ) = \,\,&\tr g( {\Yvcentermath1\,\Tiny\yng(2)\, } )\, \tr g( {\Yvcentermath1\,\Tiny\yng(1)\, } )\,\tr g( {\Yvcentermath1\,\Tiny\yng(1)\, } ) \hskip 2.5px - \tr g( {\Yvcentermath1\,\Tiny\yng(3)\, } )\,\tr g( {\Yvcentermath1\,\Tiny\yng(1)\, } )\,\tr g( {\,\boldsymbol{\cdot}\, } ) \nonumber \\
	- \,&\tr g( {\Yvcentermath1\,\Tiny\yng(2)\, } )\, \tr g( {\Yvcentermath1\,\Tiny\yng(2)\, } )\, \tr g( {\,\boldsymbol{\cdot}\, } ) + \tr g( {\Yvcentermath1\,\Tiny\yng(4)\, } )\,\tr g( {\,\boldsymbol{\cdot}\, } ) \, \tr g( {\,\boldsymbol{\cdot}\, } ). 
	\Yvcentermath0
	\label{egFs}
\end{align}
The first term on the right hand side of this equation contains the representation that is sought, along with some traces that spoil the result:
\begin{align}\Yvcentermath1
	&\tr g( {\Yvcentermath1\,\Tiny\yng(2)\, } )\, \tr g( {\Yvcentermath1\,\Tiny\yng(1)\, } )\,\tr g( {\Yvcentermath1\,\Tiny\yng(1)\, } )= \tr g( {\Yvcentermath1\,\Tiny\yng(2,1,1)\, } ) + \tr g( {\Yvcentermath1\,\Tiny\yng(2,2)\, } ) + 2\, \tr g( {\Yvcentermath1\,\Tiny\yng(3,1)\, } ) + \tr g( {\Yvcentermath1\,\Tiny\yng(4)\, } ).
	\Yvcentermath0
	\label{dec1s}
\end{align}
The last term and one copy of the second-to-last term on the right hand side of this equation are removed by the second product of traces on the top line of (\ref{egFs}) by the identity
\begin{align}\Yvcentermath1
	&\tr g( {\Yvcentermath1\,\Tiny\yng(3)\, } )\, \tr g( {\Yvcentermath1 \,\Tiny\yng(1)\, } ) = \tr g( {\Yvcentermath1\,\Tiny\yng(3,1)\, } ) + \tr g( {\Yvcentermath1\,\Tiny\yng(4)\, } ). 
	\Yvcentermath0
	\label{dec2s}
\end{align}
The second copy of the penultimate term and the trace taken in the representation with square Tableau in (\ref{dec1s}) are cancelled by the first term on the bottom line of (\ref{egFs}), but this (like in the case that the colour fields were Grassmann) also removes an extra copy of the representation whose Young Tableau has one row:
\begin{align}\Yvcentermath1
	&\tr g( {\Yvcentermath1\,\Tiny\yng(2)\, } )\, \tr g( {\Yvcentermath1 \,\Tiny\yng(2)\, } ) = \tr g( {\Yvcentermath1\,\Tiny\yng(2,2)\, } ) + \tr g( {\Yvcentermath1\,\Tiny\yng(3,1)\, } ) + \tr g( {\Yvcentermath1\,\Tiny\yng(4)\, } ). 
	\Yvcentermath0
	\label{dec3s}
\end{align}
The final term on the right hand side of (\ref{egFs}) then adds this contribution back on so as to leave only the trace of the Wilson-loop in the representation that is desired -- the first term on the right hand side of (\ref{dec1s}) with two columns in the first row and one in the next two rows. This shows how the formula (\ref{ZF''}) correctly projects onto the representation specified by the choice of Chern-Simon levels. 

For completeness we wish to point out that these techniques could now be used to write a worldline theory of a matter field in an arbitrary fully \textit{anti-}symmetric representation of the symmetry group interacting with a background field. To project onto the representation with $p$ fully anti-symmetric indices (with $0 \leq p \leq N$) requires $F = p$ families of bosonic fields. Setting $n = (1, 1, \ldots, 1)$ formula (\ref{ZF''}) provides
\begin{equation}\Yvcentermath1
	\tr g(\underset{ \,\Tiny\yng(1,1)\, }{\overset{ \,\Tiny\yng(1,1)\, } {\rvdots} })\Yvcentermath0
\end{equation}
where the Young Tableau has $p$ boxes as sought. We have checked that this is a consistent approach to take to the generation of anti-symmetric representations of the symmetry group. In particular, taking $F > N$ would imply the matter field transforming in a representation with more than $N$ fully anti-symmetrised indices. If the above technique is employed then the sum over the traces of the Wilson-loop in symmetric representations identically vanishes. The group structure which underlies this model, and the form of the $U(F)$ modular measure, ensures that this occurs. We have dealt with anti-symmetric representations fully in previous sections.

\section{Conclusion}
In this article we have presented a first quantised theory describing the Dirac field coupled with a gauge field. We represented the colour degrees of freedom of the field by introducing additional families of worldline fields which generate the required path ordering and Wilson-loop coupling. We first considered a single set of anti-commuting colour fields, whose Hilbert space contains states transforming in fully antisymmetric representations and showed that the partition of the worldline theory contains a sum over Wilson-loop interactions taken in these representations. We also described how to project onto a chosen irreducible representation by gauging a $U(1)$ symmetry on the worldline.

We then extended the theory to include $F$ copies of these colour fields which enlarge the Hilbert space to include tensor products of the representations associated to each family. To pick out a single irreducible representation from the space required the partial gauging of a $U(F)$ symmetry which rotates between these families. The effect of this gauging, and associated Chern-Simons terms, was to impose constraints on the physical states. To check that the projection we proposed in \cite{JOdof} was correct we computed the path integral over the colour fields. This produced the correct Wilson-loop coupling between the gauge field and the matter field, taken in an arbitrarily chosen representation. Our approach thus provides a complete framework for a first quantised description of arbitrary matter multiplets in the worldline formalism. 

We also repeated the above calculations in the case that the auxiliary colour fields were bosonic. The commuting nature of the colour fields then meant that our basic building blocks were Wilson-loop couplings taken in fully symmetric representations. The result of combining these objects would lead to the matter field transforming as the (reducible) tensor product of the representation provided by each family. This was overcome in the same way by partially gauging the resulting unitary symmetry and using Chern-Simons terms to fix the occupation number of each set of colour fields. We verified our construction via the path integral quantisation of the worldline theory, showing that it generates the expected Wilson-loop coupling between the matter and the gauge field, taken in the correct representation.

The work we have presented here has immediate application to scattering amplitudes in the worldline approach. In particular, it could be used to produce a Bern-Kosower type master formula for the one-loop scattering of gauge bosons in the presence of arbitrary matter. We also wish to extend our work to the case of open worldlines. In this context, the particle endpoints are fixed, the transition amplitude depends on the spin degrees of freedom at either end and the Wilson-line itself carries gauge group indices. This is important for tree level calculations and has recently been addressed for scalar field theory with one set of colour fields in \cite{NaserTree}. Along similar lines, the worldline approach to quantum field theory can be related to a theory of tensionless spinning strings which interact upon contact \cite{Us1, Us2}. The endpoints of these strings are identified as particle worldlines and the Wilson-loop interaction is contained in the partition function of the string theory. This program was initiated by a study into an Abelian-field theory but the work we have presented here would allow one to include non-Abelian interactions. This in turn would require extra worldsheet fields which are related to the colour carrying fields on the boundary and would represent interesting progress for alternative descriptions of the Dirac field.

Future work should compare and contrast the use of commuting versus anti-commuting colour fields to determine which provides the simplest and most efficient tools for calculation of physical quantities. In particular, one would not wish to compromise the existing calculational advantages offered by the worldline approach, and we hope that the method advocated in this article will preserve the ease of computation when more complicated matter is considered. One might also wish to include the colour fields in first quantised theories on curved space time, which is rapidly becoming a valuable tool to explore gravitational physics. We hope this approach will offer a powerful way of describing arbitrary matter multiplets interacting with gauge fields and be an attractive alternative to conventional tools in field theory.

\subsubsection*{Acknowledgements}
Both authors are grateful to the warm hospitality of INFN Bologna where this work was started and wish to acknowledge fruitful discussions with Fiorenzo Bastianelli and Christian Schubert that inspired much interesting progress throughout the research presented in this article. JPE wishes to thank INFN Bologna and University of Modena and Reggio Emilia for having provided financial support during the preparation of the manuscript.

\appendix
\section{Path ordered exponentials}
\label{app:A}
Here we comment on the various mechanisms by which the path ordered exponential $g(2\pi)$ is produced in the first order formalisms. We show heuristically how there are two possible prescriptions for the path ordering that arise depending upon the treatment of the colour fields and their super-particles. We consider the superspace form of the action from section~\ref{secStruct}:
\begin{equation}
	S = \int dt d\theta \, \left[-\tilde{\mathbf{\Phi}}D\mathbf{\Phi} +i\tilde{\mathbf{\Phi}}D\mathbf{X} \cdot A\left(\mathbf{X}\right)\mathbf{\Phi}\right].
\end{equation}
Gauge fixing by expanding about $e = T$ we can integrate over $\theta$ to arrive at the component form expression $S[\tilde{\phi}, \tilde{z}; \phi, z]$ in (\ref{zz}). The Green function for the free $\tilde{z}$, $z$ theory is trivial and the exponent of the path integral is linear so the (Euclidean) functional integral over these degrees of freedom is easily computed as
\begin{equation}
	\int \mathscr{D}\tilde{z}\mathscr{D}z \, e^{-S_{E}} = e^{-\int dt\, \tilde{\phi} \left(\frac{d}{dt} + \mathcal{A}^{0} - \frac{T}{2} \psi^{\mu} \left[A_{\mu}, A_{\nu}\right] \psi^{\nu}\right)\phi}
\end{equation}
so that $\mathcal{A}^{0}$ is completed to the full $\mathcal{A}$ (of course, one must then make sure that the supersymmetry variations of the remaining colour fields are then taken from (\ref{susyphi}). The Green function for the $\tilde{\phi}$, $\phi$ theory involves the step function: $G\left(t, t^{\prime}\right) \sim \Theta\left(t- t^{\prime}\right)$. This relation holds up to other terms which depend on the boundary conditions -- this is where the choice of statistics for the additional fields has an effect. As it stands, taking the Green function for the colour fields to be the step function, the integral over these fields yields a path ordering prescription
\begin{align}
		&\mathscr{P}\left(e^{i\int_{0}^{2\pi} \mathcal{A}(t) dt}\right) =  \sum_{n = 0}^{\infty} \prod_{k=0}^{n}\int_{0}^{2\pi}\!dt_{k} \prod_{i=1}^{n-1}\Theta(t_{i} - t_{i+1}) (i)^{n}{\mathcal{A}(t_{k})}
		\label{path}
\end{align}
as desired. In the main text we seek the trace of this path ordered exponentiated line integral in a chosen irreducible representation, which leads us to introduce an additional $U(1)$ field on the worldline and take into account the effect of the (anti-)periodic boundary conditions of the colour fields on the worldline Green function.

Alternatively we may form the path integral over $\tilde{\mathbf{\Phi}}$ and $\mathbf{\Phi}$ against the exponential of the action in $S$ directly. In this case it is easy to verify that $\Theta\left(t - t^{\prime} - \theta \theta^{\prime}\right)$ is a super-invariant Green function: $D\Theta\left(t - t^{\prime} - \theta \theta^{\prime}\right) = (\theta - \theta^{\prime})\delta(t - t^{\prime}) = \delta(\theta - \theta^{\prime})\delta(t - t^{\prime})$. The interaction part of the action can be expanded as
\begin{equation}
e^{-S_{\textrm{int}}} = e^{-\int dt d\theta \, \tilde{\mathbf{\Phi}}\big(  \psi \cdot A + \theta\left(\dot{\omega} \cdot A + e\psi^{\mu}\partial_{\mu}A_{\nu}\psi^{\nu}\right)\big)\mathbf{\Phi}}.
\end{equation}
Taking this as an insertion in the path integral we expand the exponential and Wick contract $\tilde{\mathbf{\Phi}}$ and $\mathbf{\Phi}$ based on the super-Green function. Using $\Theta\left(t - t^{\prime} - \theta \theta^{\prime}\right) = \Theta(t - t^{\prime}) - \theta\theta^{\prime}\delta(t - t^{\prime})$ the result can be cast into a form similar to the previous case but with a different prescription for path ordering: 
\begin{align}
		\tilde{\mathscr{P}}\left(e^{i\int_{0}^{2\pi} \theta \mathcal{A}^{0}(t) + e^{\frac{1}{2}}(t)\psi(t) \cdot A(t)\, dt}\right) &=\sum_{n = 0}^{\infty} \prod_{k=0}^{n} \int_{0}^{2\pi}\!dt_{k}d\theta_{k} \prod_{i=1}^{n-1} \left(\Theta(t_{i}\! -\! t_{i+1})\! -\!\theta_{i}\theta_{i+1}\delta(t_{i} \!-\! t_{i+1})\right)  \nonumber \\
		&\qquad\qquad\times(i)^{n}\left(\theta \mathcal{A}^{0}(t_{k}) + e^{\frac{1}{2}}\left(t_{k}\right)\psi\left(t_{k}\right) \cdot A(t_{k})\right).
\end{align}
The point is that the integral over the $\theta_{k}$ provides two terms; the order $n$ contribution containing $\mathcal{A}^{0}$ is complemented by the effect of the contact term in the super-Green function, which at order $n + 1$ multiplies two copies of the $e^{\frac{1}{2}}\psi \cdot A$ to produce the commutator $e\psi^{\mu}\left[A_{\mu}, A_{\nu}\right]\psi^{\nu}$ which completes $\mathcal{A}^{0}$ to the full $\mathcal{A}$. So the linear dependence on $A\left(\mathbf{X}\right)$ can still produce the correct super-Wilson loop exponent due to the form of the super-Green function and the result is equivalent to the familiar path ordered exponential (\ref{path}).

\section{Chern-Simons gauge fixing}
\label{app:B}
We consider the gauge fixing of the $U(1)$ theory which arose with the introduction of a projection onto states of fixed occupation number. This is of course a special case of the more general theory which requires the gauge fixing of a subgroup of a $U(F)$ worldline symmetry. At the end of this appendix we comment on this general case; whilst the general idea is the same, in the latter case we will use the Faddeev-Popov procedure to keep the discussion simple.

Under the $U(1)$ gauge transformation $\tilde{\phi} \rightarrow \tilde{\phi}e^{i\vartheta}$, $\phi \rightarrow e^{-i\vartheta}\phi$ we take $a \rightarrow a +  \dot{\vartheta}$ in order that the partition function (\ref{Za'}) be invariant. Reparameterisation invariance requires we take $a(t)$ to transform as a covariant vector (form) so that under $t \rightarrow t^{\prime}(t)$ we have $a \rightarrow \frac{dt}{dt^{\prime}}a$. This ensures that it transforms in the opposite way to the integration measure and the Chern-Simons term $\int dt \,a(t)$ is invariant. This also means that $a(t)$ transforms as $e(t)$, the square root of the worldline metric. This is why we can construct the  worldline action describing $a$ in a topologically invariant way. 

We must take care to properly account for the overcounting caused by the gauge invariance. We do so by following Polyakov \cite{polyB, Belavin}, splitting variations in $a$ into gauge transformations and orthogonal, physical changes (see also the appendix of \cite{Us2}). To define the functional integration we proceed by analogy with finite dimensional integration. If the space of fields can be parameterised by some local coordinates $\zeta_{i}$ then we define
\begin{equation}
	\mathscr{D}a = \sqrt{\det{\left(\frac{\partial a}{\partial \zeta_{i}}, \frac{\partial a}{\partial \zeta_{j}}\right)}} \prod_{k} d \zeta_{k}
	\label{Da}
\end{equation}
where $(\cdot , \cdot)$ denotes an inner product on field variations that must be chosen. We can construct such an inner product on gauge invariant variations that is reparameterisation invariant and independent of the worldline metric by making use of the transformation properties of $a$:
\begin{equation}
	\left(\delta_{1}a, \delta_{2}a\right) = \int dt \, a(t) \frac{\delta_{1}a(t) \delta_{2}a(t)}{\left(a(t)\right)^{2}}
	\label{innera}
\end{equation}
which we now use to partition the space of field variations. At the end of this section we show that, in the event that the space of functions $a(t)$ is allowed to include those where the field can vanish the same techniques can be applied by instead defining the functional integration measure using the worldline einbein. This is arguably more conventional, but it initially requires the topological independence of the theory to be broken and later restored at the end of the quantisation.

Under a gauge transformation $\delta_{1}a = \dot{\vartheta}$. For a change $\delta_{\perp}a$ to be orthogonal to this it is easy to check that we require $\frac{\delta_{\perp}a}{a} = c$, a constant, which corresponds to a global scaling of the field. Using (\ref{Da}) we then have
\begin{equation}
	\mathscr{D}a = dc \mathscr{D}\vartheta \sqrt{\left(\int dt\, a(t)\right) \det{\left(-\frac{1}{a}\frac{d}{dt} \left(\frac{1}{a}\frac{d}{dt}\right)\right)}}
	\label{Da'}
\end{equation} 
where $\mathscr{D}\vartheta$ is constructed using the reparameterisation invariant inner product on scalars:
\begin{equation}
	\left(\delta_{1}\vartheta, \delta_{2}\vartheta\right) = \int dt\, a(t) \delta_{1}\vartheta(t) \delta_{2} \vartheta(t)
\end{equation}
The gauge invariance of the action means that we may use a gauge transformation to bring an arbitrary function $a(t)$ to a constant\footnote{The finite version of the gauge transformation is $a\rightarrow a - iU^{-1}\dot{U}$, where $U = e^{i\vartheta}$. This might be used to set an arbitrary function $a(\tau)$ equal to zero but the resulting gauge transformation $U(\tau) = \exp{(-i \int_{0}^{\tau} a(\tau') d\tau')}$ is not periodic. Instead, the best than can be done is to identify a modulus of the gauge field $\theta = \int_{0}^{2\pi} a(\tau)d\tau$ and form $\tilde{U}(\tau) = U(\tau)e^{i\theta}$. The effect of this transformation is to take $a(\tau) \rightarrow \theta$ and periodicity implies $\theta \in [0, 2\pi]$. } so we use this to expand about $a(t) = \frac{\theta}{2\pi}$. Then $\int_{0}^{2\pi} dt\, a(t) = \theta$ and the eigenvalue equation for the operator in (\ref{Da'}) becomes 
\begin{equation}
	-\frac{\left(2\pi\right)^{2}}{\theta^{2}}\frac{d^{2}\Theta}{dt^{2}} = \lambda \Theta
\end{equation}
on the space of periodic functions $\Theta$. The eigenfunctions are $\sin{nt}$ and $\cos{nt}$ with two-fold degenerate eigenvalues $\lambda =  \frac{4\pi^{2}n^{2}}{\theta^{2}}$ and the determinant is therefore defined by the square of their product over the integers $n$. Using $\zeta$-function regularisation we calculate the product of eigenvalues of an operator $\hat{O}$ as
\begin{equation}
\exp{\left(-\left.\frac{d}{dz} \zeta_{\hat{O}}\left(z\right)\right|_{z = 0}\right)}\,; \quad \zeta_{\hat{O}}\left(z\right) \equiv \sum_{n = 1}^{\infty} \lambda_{n}^{-z}
\end{equation}
where the $\lambda_{n}$ are the eigenvalues of $\hat{O}$. The infinite product is then regulated via the analytic continuation of the $\zeta$-function. In our case $\zeta_{\hat{O}}(z) = \left(\frac{\theta}{2\pi}\right)^{2z} \zeta(2z)$ which has derivative
\begin{equation}
	\zeta_{\hat{O}}(z) =  2\ln{\frac{\theta}{2\pi}} \,\left(\frac{\theta}{2\pi}\right)^{2z}\zeta(2z) + 2\left(\frac{\theta}{2\pi}\right)^{2z}\zeta^{\prime}(2z).
\end{equation}
Putting $z = 0$ and exponentiating the determinant we seek is $\theta^{2}$. 

In this gauge we also have $dc = \frac{d\theta}{\theta}$ and the scalings act to translate through gauge-inequivalent fields. We must take care of the subtlety of the constant zero mode $\vartheta(t) = \varphi$ which should be excluded because it does not change the value of the gauge fixed field, so we should replace $\mathscr{D}\vartheta$ in (\ref{Da'}) by $\mathscr{D}\vartheta^{\prime}$ where $\mathscr{D}\vartheta^{\prime}$ excludes constant functions. Now the inner product between two constant functions is $\left(\delta_{1}\varphi, \delta_{2}\varphi\right) = \theta \delta_{1}\varphi \delta_{2}\varphi$. So $\mathscr{D}\vartheta = \sqrt{\theta}\mathscr{D}\vartheta'  d\varphi$. Putting all of this together we get
\begin{equation}
	\mathscr{D}a = \frac{d\theta}{2\pi} \mathscr{D}\vartheta.
\end{equation}
Now $\int \mathscr{D}\vartheta$ is just the volume of the gauge group which we have successfully separated from the integral over non-equivalent configurations. Since the action (and any insertions that may arise under the path integral) should be gauge invariant the functional integral over $\vartheta$ cancels with the overall normalisation to take into account the overcounting caused by the symmetry. In summary we can therefore choose our gauge slice to be $a(t) = \frac{\theta}{2\pi}$ and replace $\mathscr{D}a$ by $d\theta$.

The final subtlety is to notice that $a(t)$ enters in the path integral as \newline $\exp{\left(i \int_{0}^{2\pi} a(t)\, dt\right)} = e^{i\theta}$ so we must identify phases that are equivalent. To cover the space of distinguishable Wilson-loops $e^{i \theta}$ we should take $\theta \in \left[0, 2\pi\right]$ and finally arrive at the replacement
\begin{equation}
	\int \mathscr{D}a \rightarrow \int_{0}^{2\pi} \!\frac{d\theta}{2\pi}
\end{equation}
as claimed in the text.

\subsection*{Using the einbein}
It is more conventional to define reparameterisation invariant measures by making use of the worldline metric to construct expressions with the correct transformation properties. One of the benefits of doing this is that the metric is taken to be positive definite but in the case of the Chern-Simons theory introducing a dependence on the einbein affects its topological nature which we must then hope to restore at the end of the quantisation procedure. We modify the inner product (\ref{innera}) to
\begin{equation}
	\left(\delta_{1}a, \delta_{2}a\right) = \int dt \, e(t) \frac{\delta_{1}a(t) \delta_{2}a(t)}{\left(e(t)\right)^{2}}
	\label{innera2}
\end{equation}
and the inner product on scalars becomes
\begin{equation}
	\left(\delta_{1}\vartheta, \delta_{2}\vartheta\right) = \int dt\, e(t) \delta_{1}\vartheta(t) \delta_{2} \vartheta(t).
\end{equation}
We will show that the integration measure eventually comes out independent of the einbein so that this separation can be preserved despite the explicit dependence in (\ref{innera2}).

The calculation proceeds in a similar way to before. This time a variation orthogonal to a gauge transformation must satisfy $\frac{\delta_{\perp}a}{e} = c$, a constant so that the variation is proportional to the einbein itself. Then the integration measure factorises as
\begin{equation}
\mathscr{D}a = dc \mathscr{D}\vartheta \sqrt{\left(\int dt\, e(t)\right) \det{\left(-\frac{1}{e}\frac{d}{dt} \left(\frac{1}{e}\frac{d}{dt}\right)\right)}}.
	\label{Dae}
\end{equation} 
As in the main text we will fix the local reparameterisation symmetry of the worldline theory by expanding about $e = T$ and we again gauge fix the $U(1)$ symmetry by setting $a = \frac{\theta}{2\pi}$. Then $\int_{0}^{2\pi} dt \, e(t) = 2\pi T$ and the eigenvalue equation for the operator above is
\begin{equation}
	-\frac{1}{T^{2}}\frac{d^{2}\vartheta}{dt^{2}} = \lambda \vartheta,
\end{equation}
which has the same form as in the previous case. The determinant can be determined with $\zeta$-function regularisation and is given by $\left(2\pi T\right)^{2}$. This time $dc = \frac{d\theta}{2\pi T}$ and we must still take care of the zero mode in the measure $\mathscr{D\vartheta}$. The inner product between two constant functions is now $\left(\delta_{1}\varphi, \delta_{2}\varphi\right) = 2\pi T \delta_{1}\varphi \delta_{2}\varphi$, so $\mathscr{D}\vartheta = \sqrt{2\pi T}\mathscr{D}\vartheta'  d\varphi$. Combining these results we again arrive at the identification
\begin{equation}
	\int \mathscr{D}a \rightarrow \int_{0}^{2\pi} \!\frac{d\theta}{2\pi}
\end{equation}
which shows that the integration measure remains independent of $T$ after all. This concludes our detailed discussion of the gauge fixing procedure for an Abelian worldline symmetry, which is intended to illustrate the tools that could be used when this $U(1)$ symmetry is generalised to the $U(F)$ case.

\subsection*{Fixing the $U(F)$ symmetry} 
For the $U(F)$ symmetry met in sections \ref{secMixed} the gauge fixing is more complicated since the symmetry group is non-Abelian. Instead the generator of the symmetry, $\alpha$, is in the Lie algebra of $U(F)$ and effects the following gauge transformation on the gauge field:
\begin{equation}
	a \rightarrow U^{-1}aU -iU^{-1}\dot{U}; \qquad U(\tau) = e^{i\alpha(\tau)}.
\end{equation}
As before, a na\"{i}ve attempt to set the gauge field to zero requires the solution of $\frac{d}{d\tau}U(\tau) = -ia(\tau)U(\tau)$ which is easily arrived at using the path ordering prescription
\begin{equation}
	U(\tau) = \mathscr{P}e^{-i \int_{0}^{\tau} a(\tau')d\tau' }.
\end{equation}
This gauge transformation is not periodic on the circle so may not be used. Instead, we can form $\tilde{U}(\tau) = U(\tau)e^{i\Theta}$, identifying the constant group valued matrix $e^{-i\Theta} = \mathscr{P}e^{-i \int_{0}^{2\pi} a(\tau)d\tau}$. This gauge transformation takes the gauge field into the constant, upper triangular matrix $\Theta$ in the Lie algebra of the symmetry group
\begin{equation}
	a_{kj}(\tau) \rightarrow \Theta_{kj}.
\end{equation}
As discussed in \cite{JOdof}, the path integral over the colour fields is insensitive to the off diagonal terms of $a_{kj}$ so we may in fact simplify our work by choosing the gauge field to be given by the diagonal matrix $\hat{a}_{kj}$ in (\ref{afix}). Using the (Abelian) subgroup $U(1)^{F} \subset U(F)$ allows us to make a large $U(1)$ transformation on each of the $\theta_{k}$ in $\hat{a}$. Periodicity then identifies the $\theta_{k}$ as angles in $[0, 2\pi]$.

The integration measure could be defined in analogy to the $U(1)$ case considered in this appendix by defining gauge invariant and reparameterisation invariant inner products on variation in $a_{kj}$ and expanding about (\ref{afix}). However, since in the text we end up discussing the gauging of subgroups of the $U(F)$ symmetry it is more convenient to use the Faddeev-Popov formalism (see \cite{FadPop} for a nice discussion). The integration over gauge equivalent configurations can be factored out of the functional integration by integrating only over the moduli $\theta_{k}$, introducing the functional determinant (\ref{FP}) to compensate for the restriction to a chosen gauge slice. That is,
\begin{equation}
	\int \mathscr{D}a \rightarrow \int \mathscr{D}U \int \mathscr{D} a\, \Det{\left(\frac{d}{d \tau} + i[a^{U}, \cdot]\right)} \delta\left(a^{U} - \hat{a}\right)
\end{equation}
where $\hat{a}$ is the gauge fixed form of $a$ and $a^{U}$ represents the action of a gauge transformation on $a$. The calculation of these determinants is presented in the main text and, cancelling the functional integration over $U$ with the volume of the gauge group, leads to gauge fixed integrals of the form
\begin{equation}
	\int \mathscr{D} a \, \Omega[a] \rightarrow \prod_{k = 1}^{F} \int \frac{d\theta_{k}}{2\pi} \mu(\{\theta_{k}\})\, \Omega[\hat{a}(\{\theta_{k}\})]
\end{equation}
where $\Omega$ is any functional of the gauge fields and $\mu$ denotes the Faddeev-Popov measure maintaining gauge invariance. 
\newpage
\bibliographystyle{JHEP}
\bibliography{bibSym-v2}

\providecommand{\href}[2]{#2}\begingroup\raggedright\begin{thebibliography}{10}

\bibitem{Strass1}
M.~J. Strassler, \emph{{Field theory without Feynman diagrams: One loop
  effective actions}},
  \href{http://dx.doi.org/10.1016/0550-3213(92)90098-V}{\emph{Nucl. Phys.} {\bf
  B385} (1992) 145--184}, [\href{http://arxiv.org/abs/hep-ph/9205205}{{\tt
  hep-ph/9205205}}].

\bibitem{Schu}
C.~Schubert, \emph{{Perturbative quantum field theory in the string inspired
  formalism}},
  \href{http://dx.doi.org/10.1016/S0370-1573(01)00013-8}{\emph{Phys. Rept.}
  {\bf 355} (2001) 73--234}, [\href{http://arxiv.org/abs/hep-th/0101036}{{\tt
  hep-th/0101036}}].

\bibitem{Bern}
Z.~Bern and D.~A. Kosower, \emph{{Efficient calculation of one loop QCD
  amplitudes}},
  \href{http://dx.doi.org/10.1103/PhysRevLett.66.1669}{\emph{Phys. Rev. Lett.}
  {\bf 66} (1991) 1669--1672}.

\bibitem{Sato}
H.-T. Sato and M.~G. Schmidt, \emph{{Worldline approach to the Bern-Kosower
  formalism in two loop Yang-Mills theory}},
  \href{http://dx.doi.org/10.1016/S0550-3213(99)00386-7}{\emph{Nucl.Phys.} {\bf
  B560} (1999) 551--586}, [\href{http://arxiv.org/abs/hep-th/9812229}{{\tt
  hep-th/9812229}}].

\bibitem{NaserTree}
N.~Ahmadiniaz, F.~Bastianelli and O.~Corradini, \emph{{Dressed scalar
  propagator in a non-abelian background from the worldline formalism}},
  \href{http://dx.doi.org/10.1103/PhysRevD.93.025035}{\emph{Phys. Rev.} {\bf
  D93} (2016) 025035}, [\href{http://arxiv.org/abs/1508.05144}{{\tt
  1508.05144}}].

\bibitem{Schmidt}
M.~G. Schmidt and C.~Schubert, \emph{{On the calculation of effective actions
  by string methods}},
  \href{http://dx.doi.org/10.1016/0370-2693(93)91537-W}{\emph{Phys.Lett.} {\bf
  B318} (1993) 438--446}, [\href{http://arxiv.org/abs/hep-th/9309055}{{\tt
  hep-th/9309055}}].

\bibitem{Dunne}
G.~V. Dunne and C.~Schubert, \emph{{Multiloop information from the QED
  effective Lagrangian}},
  \href{http://dx.doi.org/10.1088/1742-6596/37/1/012}{\emph{J.Phys.Conf.Ser.}
  {\bf 37} (2006) 59--72}, [\href{http://arxiv.org/abs/hep-th/0409021}{{\tt
  hep-th/0409021}}].

\bibitem{Bgrav1}
F.~Bastianelli, J.~M. Davila and C.~Schubert, \emph{{Gravitational corrections
  to the Euler-Heisenberg Lagrangian}},
  \href{http://dx.doi.org/10.1088/1126-6708/2009/03/086}{\emph{JHEP} {\bf 0903}
  (2009) 086}, [\href{http://arxiv.org/abs/0812.4849}{{\tt 0812.4849}}].

\bibitem{hspin2}
F.~Bastianelli, O.~Corradini and E.~Latini, \emph{{Higher spin fields from a
  worldline perspective}},
  \href{http://dx.doi.org/10.1088/1126-6708/2007/02/072}{\emph{JHEP} {\bf 02}
  (2007) 072}, [\href{http://arxiv.org/abs/hep-th/0701055}{{\tt
  hep-th/0701055}}].

\bibitem{Anton}
A.~Ilderton, \emph{{Localisation in worldline pair production and lightfront
  zero-modes}}, \href{http://dx.doi.org/10.1007/JHEP09(2014)166}{\emph{JHEP}
  {\bf 1409} (2014) 166}, [\href{http://arxiv.org/abs/1406.1513}{{\tt
  1406.1513}}].

\bibitem{Bastbrst}
F.~Bastianelli, O.~Corradini and A.~Waldron, \emph{{Detours and Paths: BRST
  Complexes and Worldline Formalism}},
  \href{http://dx.doi.org/10.1088/1126-6708/2009/05/017}{\emph{JHEP} {\bf 05}
  (2009) 017}, [\href{http://arxiv.org/abs/0902.0530}{{\tt 0902.0530}}].

\bibitem{Reuter}
M.~Reuter, M.~G. Schmidt and C.~Schubert, \emph{{Constant external fields in
  gauge theory and the spin 0, 1/2, 1 path integrals}},
  \href{http://dx.doi.org/10.1006/aphy.1997.5716}{\emph{Annals Phys.} {\bf 259}
  (1997) 313--365}, [\href{http://arxiv.org/abs/hep-th/9610191}{{\tt
  hep-th/9610191}}].

\bibitem{Mond}
M.~Mondragon, L.~Nellen, M.~G. Schmidt and C.~Schubert, \emph{{Axial couplings
  on the worldline}},
  \href{http://dx.doi.org/10.1016/0370-2693(95)01392-X}{\emph{Phys. Lett.} {\bf
  B366} (1996) 212--219}, [\href{http://arxiv.org/abs/hep-th/9510036}{{\tt
  hep-th/9510036}}].

\bibitem{Bastianelli:2008vh}
F.~Bastianelli, O.~Corradini, P.~A.~G. Pisani and C.~Schubert, \emph{{Scalar
  heat kernel with boundary in the worldline formalism}},
  \href{http://dx.doi.org/10.1088/1126-6708/2008/10/095}{\emph{JHEP} {\bf 10}
  (2008) 095}, [\href{http://arxiv.org/abs/0809.0652}{{\tt 0809.0652}}].

\bibitem{Bastianelli:2000dw}
F.~Bastianelli and O.~Corradini, \emph{{6-D trace anomalies from quantum
  mechanical path integrals}},
  \href{http://dx.doi.org/10.1103/PhysRevD.63.065005}{\emph{Phys. Rev.} {\bf
  D63} (2001) 065005}, [\href{http://arxiv.org/abs/hep-th/0010118}{{\tt
  hep-th/0010118}}].

\bibitem{grass}
A.~Balachandran, P.~Salomonson, B.-S. Skagerstam and J.-O. Winnberg,
  \emph{{Classical Description of Particle Interacting with Nonabelian Gauge
  Field}}, \href{http://dx.doi.org/10.1103/PhysRevD.15.2308}{\emph{Phys.Rev.}
  {\bf D15} (1977) 2308}.

\bibitem{grass2}
A.~Barducci, R.~Casalbuoni and L.~Lusanna, \emph{{Classical Scalar and Spinning
  Particles Interacting with External Yang-Mills Fields}},
  \href{http://dx.doi.org/10.1016/0550-3213(77)90278-4}{\emph{Nucl. Phys.} {\bf
  B124} (1977) 93}.

\bibitem{Paul2}
P.~Mansfield, \emph{{The fermion content of the Standard Model from a simple
  world-line theory}},
  \href{http://dx.doi.org/10.1016/j.physletb.2015.02.061}{\emph{Phys. Lett.}
  {\bf B743} (2015) 353--356}, [\href{http://arxiv.org/abs/1410.7298}{{\tt
  1410.7298}}].

\bibitem{Me1}
J.~P. Edwards, \emph{{Unified theory in the worldline approach}},
  \href{http://dx.doi.org/10.1016/j.physletb.2015.09.038}{\emph{Phys. Lett.}
  {\bf B750} (2015) 312--318}, [\href{http://arxiv.org/abs/1411.6540}{{\tt
  1411.6540}}].

\bibitem{Bastvt}
F.~Bastianelli, P.~Benincasa and S.~Giombi, \emph{{Worldline approach to vector
  and antisymmetric tensor fields}},
  \href{http://dx.doi.org/10.1088/1126-6708/2005/04/010}{\emph{JHEP} {\bf 0504}
  (2005) 010}, [\href{http://arxiv.org/abs/hep-th/0503155}{{\tt
  hep-th/0503155}}].

\bibitem{Bastforms2}
F.~Bastianelli and R.~Bonezzi, \emph{{Quantum theory of massless (p,0)-forms}},
  \href{http://dx.doi.org/10.1007/JHEP09(2011)018}{\emph{JHEP} {\bf 09} (2011)
  018}, [\href{http://arxiv.org/abs/1107.3661}{{\tt 1107.3661}}].

\bibitem{JOdof}
O.~Corradini and J.~P. Edwards, \emph{{Mixed symmetry tensors in the worldline
  formalism}}, \href{http://dx.doi.org/10.1007/JHEP05(2016)056}{\emph{JHEP}
  {\bf 05} (2016) 056}, [\href{http://arxiv.org/abs/1603.07929}{{\tt
  1603.07929}}].

\bibitem{Bastwl1}
F.~Bastianelli, R.~Bonezzi, O.~Corradini and E.~Latini, \emph{{Particles with
  non abelian charges}},
  \href{http://dx.doi.org/10.1007/JHEP10(2013)098}{\emph{JHEP} {\bf 1310}
  (2013) 098}, [\href{http://arxiv.org/abs/1309.1608}{{\tt 1309.1608}}].

\bibitem{BdVH}
L.~Brink, P.~Di~Vecchia and P.~Howe, \emph{A {L}agrangian formulation of the
  classical and quantum dynamics of spinning particles}, {\emph{Nucl. Phys. B}
  {\bf 118} (1977) 76--94}.

\bibitem{Strass2}
M.~J. Strassler, \emph{Field theory without feynman diagrams: A demonstration
  using actions induced by heavy particles}, {\emph{SLAC-PUB-5978} (1992) }.

\bibitem{Schwinger}
J.~Schwinger, \emph{On gauge invariance and vacuum polarization},
  \href{http://dx.doi.org/10.1103/PhysRev.82.664}{\emph{Phys. Rev.} {\bf 82}
  (Jun, 1951) 664--679}.

\bibitem{Sam}
S.~Samuel, \emph{Color zitterbewegung}, {\emph{Nucl. Phys.} {\bf B149} (1979)
  517}.

\bibitem{Bastwl2}
F.~Bastianelli, R.~Bonezzi, O.~Corradini, E.~Latini and K.~H. Ould-Lahoucine,
  \emph{{A worldline approach to colored particles}},
  \href{http://arxiv.org/abs/1504.03617}{{\tt 1504.03617}}.

\bibitem{hspin}
F.~Bastianelli, R.~Bonezzi, O.~Corradini and E.~Latini, \emph{{Spinning
  particles and higher spin field equations}},
  \href{http://arxiv.org/abs/1504.02683}{{\tt 1504.02683}}.

\bibitem{hspin3}
F.~Bastianelli, O.~Corradini and E.~Latini, \emph{{Spinning particles and
  higher spin fields on (A)dS backgrounds}},
  \href{http://dx.doi.org/10.1088/1126-6708/2008/11/054}{\emph{JHEP} {\bf 11}
  (2008) 054}, [\href{http://arxiv.org/abs/0810.0188}{{\tt 0810.0188}}].

\bibitem{Corradini:2010ia}
O.~Corradini, \emph{{Half-integer Higher Spin Fields in (A)dS from Spinning
  Particle Models}},
  \href{http://dx.doi.org/10.1007/JHEP09(2010)113}{\emph{JHEP} {\bf 1009}
  (2010) 113}, [\href{http://arxiv.org/abs/1006.4452}{{\tt 1006.4452}}].

\bibitem{Aux}
O.~Andreev and A.~Tseytlin, \emph{Generating functional for scattering
  amplitudes and the effective action in the open superstring theory},
  \href{http://dx.doi.org/http://dx.doi.org/10.1016/0370-2693(88)91408-6}{\emph{Physics
  Letters B} {\bf 207} (1988) 157 -- 163}.

\bibitem{West}
P.~West, \emph{Introduction to supersymmetry and supergravity}.
\newblock No.~25. World Scientific, 2~ed., 1990.

\bibitem{Howe}
P.~Howe, S.~Penati, M.~Pernici and P.~K. Townsend, \emph{A particle mechanics
  description of antisymmetric tensor fields}, {\emph{Classical and Quantum
  Gravity} {\bf 6} (1989) 1125}.

\bibitem{Hoker}
E.~D'Hoker and D.~G. Gagne, \emph{{Worldline path integrals for fermions with
  general couplings}},
  \href{http://dx.doi.org/10.1016/0550-3213(96)00126-5}{\emph{Nucl.Phys.} {\bf
  B467} (1996) 297--312}, [\href{http://arxiv.org/abs/hep-th/9512080}{{\tt
  hep-th/9512080}}].

\bibitem{anom}
S.~Elitzur, E.~Rabinovici, Y.~Frishman and A.~Schwimmer, \emph{Origins of
  global anomalies in quantum mechanics},
  \href{http://dx.doi.org/10.1016/0550-3213(86)90042-8}{\emph{Nuclear Physics
  B} {\bf 273} (1986) 93 -- 108}.

\bibitem{Dai}
P.~Dai, Y.-t. Huang and W.~Siegel, \emph{{Worldgraph Approach to Yang-Mills
  Amplitudes from N=2 Spinning Particle}},
  \href{http://dx.doi.org/10.1088/1126-6708/2008/10/027}{\emph{JHEP} {\bf 0810}
  (2008) 027}, [\href{http://arxiv.org/abs/0807.0391}{{\tt 0807.0391}}].

\bibitem{Bastforms}
F.~Bastianelli, R.~Bonezzi and C.~Iazeolla, \emph{{Quantum theories of
  (p,q)-forms}}, \href{http://dx.doi.org/10.1007/JHEP08(2012)045}{\emph{JHEP}
  {\bf 1208} (2012) 045}, [\href{http://arxiv.org/abs/1204.5954}{{\tt
  1204.5954}}].

\bibitem{Us2}
J.~P. Edwards and P.~Mansfield, \emph{{QED as the tensionless limit of the
  spinning string with contact interaction}},
  \href{http://dx.doi.org/10.1016/j.physletb.2015.05.024}{\emph{Phys. Lett.}
  {\bf B746} (2015) 335--340}, [\href{http://arxiv.org/abs/1409.4948}{{\tt
  1409.4948}}].

\bibitem{Bastianelli:2011cc}
F.~Bastianelli, R.~Bonezzi, O.~Corradini and E.~Latini, \emph{{Extended SUSY
  quantum mechanics: transition amplitudes and path integrals}},
  \href{http://dx.doi.org/10.1007/JHEP06(2011)023}{\emph{JHEP} {\bf 06} (2011)
  023}, [\href{http://arxiv.org/abs/1103.3993}{{\tt 1103.3993}}].

\bibitem{Bastianelli:2012bn}
F.~Bastianelli, R.~Bonezzi, O.~Corradini and E.~Latini, \emph{{Effective action
  for higher spin fields on (A)dS backgrounds}},
  \href{http://dx.doi.org/10.1007/JHEP12(2012)113}{\emph{JHEP} {\bf 12} (2012)
  113}, [\href{http://arxiv.org/abs/1210.4649}{{\tt 1210.4649}}].

\bibitem{Bastianelli:2014lia}
F.~Bastianelli, R.~Bonezzi, O.~Corradini and E.~Latini, \emph{{Massive and
  massless higher spinning particles in odd dimensions}},
  \href{http://dx.doi.org/10.1007/JHEP09(2014)158}{\emph{JHEP} {\bf 09} (2014)
  158}, [\href{http://arxiv.org/abs/1407.4950}{{\tt 1407.4950}}].

\bibitem{Hawk}
S.~Hawking, \emph{{Zeta Function Regularization of Path Integrals in Curved
  Space-Time}},
  \href{http://dx.doi.org/10.1007/BF01626516}{\emph{Commun.Math.Phys.} {\bf 55}
  (1977) 133}.

\bibitem{Me2un}
J.~P. Edwards, \emph{{Contact interactions between particle worldlines}},
  \href{http://dx.doi.org/10.1007/JHEP01(2016)033}{\emph{JHEP} {\bf 01} (2016)
  033}, [\href{http://arxiv.org/abs/1506.08130}{{\tt 1506.08130}}].

\bibitem{Me2Th}
J.~P. Edwards, \emph{Contact interactions for point particles and strings}.
\newblock PhD thesis, University of Durham, 2015.

\bibitem{Bastforms3}
F.~Bastianelli and R.~Bonezzi, \emph{{U(N) spinning particles and higher spin
  equations on complex manifolds}},
  \href{http://dx.doi.org/10.1088/1126-6708/2009/03/063}{\emph{JHEP} {\bf 03}
  (2009) 063}, [\href{http://arxiv.org/abs/0901.2311}{{\tt 0901.2311}}].

\bibitem{Us1}
J.~P. Edwards and P.~Mansfield, \emph{{Delta-function Interactions for the
  Bosonic and Spinning Strings and the Generation of Abelian Gauge Theory}},
  \href{http://dx.doi.org/10.1007/JHEP01(2015)127}{\emph{JHEP} {\bf 1501}
  (2015) 127}, [\href{http://arxiv.org/abs/1410.3288}{{\tt 1410.3288}}].

\bibitem{polyB}
A.~M. Polyakov, \emph{Quantum geometry of bosonic strings}, {\emph{Phys. Let.
  B} (1981) 207--210}.

\bibitem{Belavin}
A.~Belavin and G.~Tarnopolsky, \emph{{Introduction to string theory and
  conformal field theory}},
  \href{http://dx.doi.org/10.1134/S1063778810050108}{\emph{Phys.Atom.Nucl.}
  {\bf 73} (2010) 848--877}.

\bibitem{FadPop}
M.~Ornigotti and A.~Aiello, \emph{{The Faddeev-Popov Method Demystified}},
  \href{http://arxiv.org/abs/1407.7256}{{\tt 1407.7256}}.

\end{thebibliography}\endgroup
\end{document}